\documentclass[11pt]{article}

\usepackage{setspace}

\usepackage{fullpage,float}
\usepackage{atmacros}
\usepackage[T1]{fontenc}
\usepackage{gitinfo}
\usepackage[usenames,dvipsnames]{xcolor}
\usepackage[shortlabels]{enumitem}
\usepackage{stmaryrd}
\usepackage{mathtools}

\usepackage{mathpazo}
\usepackage{bm}
\usepackage{todonotes}
\usepackage{lipsum}
\usepackage{scrextend}
\usepackage{xfrac}

\newcommand{\TR}{{\sf TR}}

\usepackage[linesnumbered,vlined,boxed,ruled]{algorithm2e}
\makeatletter
\newcommand{\RemoveAlgoNumber}{\renewcommand{\fnum@algocf}{\AlCapSty{\AlCapFnt\algorithmcfname}}}
\newcommand{\RevertAlgoNumber}{\algocf@resetfnum}
\makeatother

\RequirePackage[colorlinks=true,backref=page]{hyperref}
\hypersetup{
  linkcolor=[rgb]{0.3,0.3,0.6},
  citecolor=[rgb]{0.2, 0.6, 0.2},
  urlcolor=[rgb]{0.6, 0.2, 0.2}
}

\usepackage{nth}
\usepackage{intcalc}
\usepackage{etoolbox}
\usepackage{xstring}

\usepackage{ifpdf}
\ifpdf
\else
\usepackage[quadpoints=false]{hypdvips}
\fi

\newcommand{\ECCCurl}[2]{http://eccc.hpi-web.de/report/\ifnumcomp{#1}{>}{93}{19}{20}#1/#2/}

\newcommand{\shortECCC}[2]{\texttt{\href{http://eccc.hpi-web.de/report/\ifnumcomp{#1}{>}{93}{19}{20}#1/#2/}{eccc:TR#1-#2}}}

\newcommand{\parseECCC}[1]{
\StrSubstitute{#1}{TR}{}[\tmpstring]%
\IfSubStr{\tmpstring}{/}{ 
\StrBefore{\tmpstring}{/}[\ecccyear]%
\StrBehind{\tmpstring}{/}[\ecccreport]%
}{
\StrBefore{\tmpstring}{-}[\ecccyear]%
\StrBehind{\tmpstring}{-}[\ecccreport]%
}%
\shortECCC{\ecccyear}{\ecccreport}}

\usepackage{amsthm}
\usepackage{thmtools,thm-restate}

\numberwithin{equation}{section}
\declaretheoremstyle[bodyfont=\it,qed=\qedsymbol]{noproofstyle}

\declaretheorem[numberlike=equation]{observation}

\declaretheorem[name=Observation,numbered=no]{observation*}

\declaretheorem[numberlike=equation]{fact}

\declaretheorem[numberlike=equation]{theorem}

\declaretheorem[name=Theorem,numbered=no]{theorem*}

\declaretheorem[numberlike=equation]{lemma}
\declaretheorem[name=Lemma,numbered=no]{lemma*}

\declaretheorem[numberlike=equation]{corollary}
\declaretheorem[name=Corollary,numbered=no]{corollary*}

\declaretheorem[name=Proposition,numbered=no]{proposition*}

\declaretheorem[name=Claim,numbered=no]{claim*}

\declaretheorem[numberlike=equation]{conjecture}
\declaretheorem[name=Conjecture,numbered=no]{conjecture*}

\declaretheorem[numberlike=equation]{question}
\declaretheorem[name=Question,numbered=no]{question*}

\declaretheoremstyle[bodyfont=\it,qed=$\lozenge$]{defstyle} 

\declaretheorem[numberlike=equation,style=defstyle]{definition}
\declaretheorem[unnumbered,name=Definition,style=defstyle]{definition*}

\declaretheorem[numberlike=equation,style=defstyle]{example}
\declaretheorem[unnumbered,name=Example,style=defstyle]{example*}

\declaretheorem[unnumbered,name=Notation=defstyle]{notation*}

\declaretheorem[numberlike=equation,style=defstyle]{construction}
\declaretheorem[unnumbered,name=Construction,style=defstyle]{construction*}

\declaretheorem[numberlike=equation,style=defstyle]{remark}
\declaretheorem[unnumbered,name=Remark,style=defstyle]{remark*}

\newcommand{\ehref}[1]{\href{mailto:#1}{#1}}
\newcommand{\ignore}[1]{}

\newcommand{\inparen}[1]{\left( #1 \right)}

\newcommand{\inbrace}[1]{\left\{ #1 \right\}}

\newcommand{\inangle}[1]{\left< #1 \right>}

\newcommand{\set}[1]{\inbrace{#1}}

\newcommand{\abs}[1]{\left| #1 \right|}

\newcommand{\coeff}{\operatorname{coeff}}

\newcommand{\supp}{\operatorname{supp}}

\newcommand{\Fspan}{\operatorname{span}}

\def\VP{\mathsf{VP}}
\def\VNP{\mathsf{VNP}}

\def\P{\mathsf{P}}

\def\NP{\mathsf{NP}}

\newcommand{\VBP}{\textsf{VBP}}

\newcommand{\calc}{\mathcal{C}}

\newcommand{\C}{\mathbb{C}}

\newcommand{\ESym}{\operatorname{\sf ESym}}

\newcommand{\ROABP}{\operatorname{ROABP}}

\newcommand{\commROABP}{\mathsf{commROABP}}
\newcommand{\diagROABP}{\mathsf{diagROABP}}

\newcommand{\waring}{\operatorname{WR}}
\newcommand{\SES}{\Sigma\bigwedge\Sigma}

\newcommand{\V}{\mathbf{V}}

\newcommand{\normalset}{\operatorname{NS}}

\newcommand{\dotProd}[2]{\inangle{#1,#2}}

\newcommand{\trans}[1]{{#1}^{\intercal}}
\newcommand{\ideal}[1]{\inangle{#1}}
\newcommand{\roots}{\operatorname{roots}}

\newcommand{\ltmon}{\prec}
\newcommand{\LM}{\operatorname{LM}}
\newcommand{\partials}{\operatorname{DPD}}

\newcommand{\vecA}{\mathbf{A}}

\newif\ifdraft
\drafttrue 

\ifdraft

\newcommand{\ATnote}[1]{\textcolor{OliveGreen}{\guillemotleft AT: #1 \guillemotright}}
\newcommand{\CRnote}[1]{\textcolor{Purple}{\guillemotleft CR: #1 \guillemotright}}
\newcommand{\pending}[1]{#1 \vspace{1em}}
\else

\newcommand{\ATnote}[1]{}
\newcommand{\CRnote}[1]{}
\newcommand{\pending}[1]{}
\fi

\title{On Finer Separations between Subclasses of\\Read-once Oblivious ABPs}

\author{
  C. Ramya\thanks{\ehref{c.ramya@cmi.ac.in}, Chennai Mathematical Institute, India. Research supported by INSPIRE Faculty Fellowship of DST and by a grant from the Infosys Foundation. Part of this work was done when at the Tata Institute of Fundamental Research, Mumbai, India (DAE project 12-R\&D-TFR-5.01-0500).}
  \and
  Anamay Tengse\thanks{\ehref{anamay.tengse@gmail.com}, Dept. of Computer Science, University of Haifa, Israel. Research supported by the Israel Science Foundation (grant No. 716/20). Part of this work was done when at the Tata Institute of Fundamental Research, Mumbai, India, as a student (DAE project no 12-R\&D-TFR-5.01-0500) and as a visitor (Prof. Prahladh Harsha's Swarnajaynti fellowship and Prof. Arkadev Chattopadhyay's Microsoft Research funds).}
}

\begin{document}

\maketitle

{\let\thefootnote\relax
\footnotetext{\textcolor{white}{(Fun fact: `Timbuktu' is an actual city in Mali!) Base version:~(\gitAuthorIsoDate)\;,\;\gitAbbrevHash\;\; \gitVtag}}
}

\begin{abstract}
Read-once Oblivious Algebraic Branching Programs (ROABPs) compute polynomials as products of univariate polynomials that have matrices as coefficients. In an attempt to understand the landscape of algebraic complexity classes surrounding ROABPs, we study classes of ROABPs based on the algebraic structure of these coefficient matrices. We study connections between polynomials computed by these structured variants of ROABPs and other well-known classes of polynomials (such as depth-three powering circuits, tensor-rank and Waring rank of polynomials).

Our main result concerns \emph{commutative ROABPs}, where \emph{all} coefficient matrices commute with each other, and \emph{diagonal ROABPs}, where all the coefficient matrices are just diagonal matrices. In particular, we show a somewhat surprising connection between these models and the model of \emph{depth-three powering circuits} that is related to the \emph{Waring rank} of polynomials. We show that if the \emph{dimension of partial derivatives} captures \emph{Waring rank} up to polynomial factors, then the model of \emph{diagonal ROABPs} efficiently simulates the seemingly more expressive model of \emph{commutative ROABPs}.
Further, a \emph{commutative ROABP} that cannot be efficiently simulated by a \emph{diagonal ROABP} will give an explicit polynomial that gives a super-polynomial separation between \emph{dimension of partial derivatives} and \emph{Waring rank}.

Our proof of the above result builds on the results of Marinari, {\Moller} and Mora (1993), and {\Moller} and Stetter (1995), that characterise rings of commuting matrices in terms of polynomials that have small dimension of partial derivatives. The algebraic structure of the coefficient matrices of these ROABPs plays a crucial role in our proofs.
\end{abstract}

\newpage


\section{Introduction}
\label{sec:introduction}

The central question in \emph{algebraic complexity theory}: the theory concerning computation of polynomials, is to understand the most efficient way of computing a polynomial $f(x_1,\ldots,x_n)$ using the basic arithmetic operations of addition and multiplication.
One of the earliest works to study the computational complexity of an \emph{explicit} polynomial is perhaps the famous work of Strassen~\cite{S69} on matrix multiplication.
However, the seminal work of Valiant~\cite{V79} that proposed the ``$\VP$ vs $\VNP$'' question (the algebraic analogue of $\P$ vs $\NP$) is widely regarded as the starting point of algebraic complexity theory.

\emph{Algebraic circuits} are a fundamental model for computing polynomials, and the complexity of a polynomial is determined by the \emph{size} of the smallest circuit that computes it.
This definition also coincides with the fewest number of arithmetic operations required to evaluate a polynomial.
Valiant’s above mentioned work however, uses the model of \emph{algebraic branching programs (ABPs)} to capture \emph{efficiently computable polynomials}.
Informally, an ABP computes a polynomial $f(\vecx)$ as the $(1,1)$th entry of a product of matrices, each of which has linear forms in the $\vecx$ variables as its entries.
While $\VP$ is the class of $n$-variate polynomials having $\poly(n)$ size algebraic circuits, the class of $n$-variate polynomials that have an ABP of size $\poly(n)$ is called $\VBP$.
The class $\VBP$ is known to be a subclass of $\VP$, and at the moment it is unclear if this inclusion is strict.
The $\VBP$ vs $\VNP$ question remains a central question in algebraic complexity theory as it is captured by the ``determinant vs permanent'' question (see e.g. \cite{KV21}).

Although proving strong lower bounds against algebraic circuits seems currently unattainable, even proving lower bounds against ABPs remains a challenging task.
In fact, even a super-quadratic lower bound against ABPs will be a massive improvement over the state of the art (\cite{BS83, CKSV20}).
A significant amount of work in the area has therefore focused on analysing more structured variants of ABPs which could potentially be easier to tackle. Indeed, a celebrated result of Nisan \cite{N91} gives an exact characterisation of the complexity of a \emph{non-commutative ABP} computing any non-commutative polynomial\footnote{A non-commutative polynomial is one in which the variables do not commute, i.e. $x y \neq y x$.}.
This characterisation yields a $2^{\Omega(n)}$ lower bound against non-commutative ABPs for the determinant, which among other things, highlights the power of commutativity.

We now turn to the protagonists of our work, Read-once Oblivious ABPs (ROABPs), which are the commutative analogues of non-commutative ABPs.
ROABPs were first introduced by Forbes and Shpilka~\cite{FS13}, in the context of \emph{polynomial identity testing}: another central problem in algebraic complexity, which we discuss in more detail in \autoref{sec:pit}.
An ROABP is an algebraic branching program that uses exactly $n$ matrices, one for each variable; and the entries in the matrix corresponding to an $x_i $ are univariate polynomials from $\C[x_i]$ (formally defined in \autoref{defn:roabp}).
It is easy to check that ROABPs can compute any monomial, and are closed under taking sums.
Thus, every $n$-variate, degree-$d$ polynomial trivially has an ROABP of size $d^{O(n)}$.
On the other hand, Nisan's characterisation~\cite{N91} for non-commutative ABPs also extends to ROABPs, and hence most of the strong lower bounds against non-commutative ABPs can be suitably translated to ROABPs. 

Since all ROABPs use $n$ matrices, the parameter of interest is the \emph{width} of an ROABP, which is the maximum dimension of any of the underlying matrices.
Furthermore, since every matrix in an ROABP is associated with exactly one variable in $\{x_1,\ldots ,x_n\}$, one can naturally identify an order $\sigma \in S_n $ (permutation on $\{x_1,\ldots ,x_n\}$) in which the ROABP ``reads the variables''.
Indeed, there are polynomials which are computable by $\poly(n)$-width ROABPs in one order, but require exponential width in a different order.
In fact, a straight-forward application of Nisan's characterisation shows that the $2n$-variate polynomial $(x_1+y_1)(x_2+y_2)\cdots (x_n+y_n)$ is computable by a width-$2$ ROABP in the order $(x_1,y_1,x_2,y_2,\ldots ,x_n,y_n)$; but any ROABP that reads all the $\vecx$-variables before the $\vecy$-variables (e.g. in the order $(x_1,\ldots ,x_n,y_1,\ldots ,y_n)$) requires width $2^{\Omega(n)}$.
The existence of such polynomials naturally leads to the following classes of polynomials (defined in \autoref{sec:preliminaries}).
\begin{itemize}\itemsep 0pt
\item $\ROABP[\exists](n,d,w)$ - $n$-variate, individual degree $d$ polynomials that are computable by a width-$w$ ROABP in \emph{some} order $\sigma \in S_n$. 
\item $\ROABP[\forall](n,d,w)$ - $n$-variate, individual degree $d$ polynomials that are computable by a width-$w$ ROABP in \emph{every} order.
\end{itemize}
Clearly, $\ROABP[\forall](n,d,w) \subseteq \ROABP[\exists](n,d,w)$, and the former class requires exponential width to simulate the latter, due to the example discussed above.

Observe that an ROABP in the order $\mathrm{id} = (x_1,\ldots,x_n)$, can be written as $\trans{\vecu} \cdot \inparen{\prod_{i \in [n]} M_i(x_i)} \cdot \vecv$, with entries of each $M_i$ being univariate polynomials in $\mathbb{C}[x_i]$.
Alternatively, we can view the same, as $\trans{\vecu} \inparen{\prod_{i \in [n]} \inparen{ A_{i,0} + A_{i,1} x_i + \cdots + A_{i,d} x_i^d } } \vecv$, by interpreting each $M_i$ as a univariate with matrices as coefficients.
We refer to these matrices $\set{A_{i,j}}$ as the \emph{coefficient matrices} of the ROABP. 

Now based on the properties of the coefficient matrices $\set{A_{i,j}}$, one can define the following models and the corresponding classes.
\begin{itemize}
\item \emph{Commutative ROABPs:} ROABPs where all the $n(d+1)$ coefficient matrices commute with each other (see \autoref{defn:commutative-ROABP}).\\
$\commROABP(n,d,w)$ - $n$-variate, individual degree $d$ polynomials that are computable by a width $w$ commutative ROABP.
\item \emph{Diagonal ROABPs:} ROABPs where all the $n(d+1)$ coefficient matrices are diagonal matrices (see \autoref{defn:diagonal-ROABP}).\\
$\diagROABP(n,d,w)$ - $n$-variate, individual degree $d$ polynomials that are computable by a width $w$ diagonal ROABP.
\end{itemize}

First of all, $\commROABP(n,d,w) \subseteq \ROABP[\forall](n,d,w)$ for any $n,d,w$, since the coefficient matrices in any commutative ROABP are commutative, and one can multiply the matrices in any order to get the same result.
Likewise, as all diagonal matrices commute with each other, $\diagROABP(n,d,w) \subseteq \commROABP(n,d,w)$.
In this paper, we investigate commutative and diagonal ROABPs to understand if and when these two classes are the essentially (up to polynomial-factors) equal.

While it is indeed true that even diagonal ROABPs are universal, it is reasonable to ask if there are any interesting polynomial families that are efficiently computable by commutative and diagonal ROABPs.
In this regard, let us begin by looking at the constructions of ``all-order-ROABPs'' for two well studied polynomial families: \emph{elementary symmetric polynomials} and \emph{powers of linear forms}.
Incidentally, these constructions can naturally be interpreted as commutative ROABPs, and further, they even lead to diagonal ROABPs that achieve the best known upper bounds.
We believe that these examples should serve as an additional motivation to study the models of commutative and diagonal ROABPs.

\begin{figure}[ht]
\begin{center}
\begin{tikzpicture}[scale=0.8]

  \node[draw, circle] (s) at (-6,0) {$s$};

  \draw[fill=black] (-4,3) circle (1.5pt)
  edge [<-] (s);
  \draw[fill=black] (-4,1) circle (1.5pt)
  edge [<-, color=gray] node[above] {$0$} (s);
  \draw[fill=black] (-4,-1) circle (1.5pt)
  edge [<-, color=gray] node[above] {$0$} (s);
  \draw[fill=black] (-4,-3) circle (1.5pt)
  edge [<-, color=gray] node[above] {$0$} (s);

  \draw[fill=black] (-2,3) circle (1.5pt)
  edge [<-] (-4,3);
  \draw[fill=black] (-2,1) circle (1.5pt)
  edge [<-] (-4,1)
  edge [<-] node[above] {$x_1$} (-4,3);
  \draw[fill=black] (-2,-1) circle (1.5pt)
  edge [<-] (-4,-1)
  edge [<-] node[above] {$x_1$} (-4,1);
  \draw[fill=black] (-2,-3) circle (1.5pt)
  edge [<-] (-4,-3)
  edge [<-] node[above] {$x_1$} (-4,-1);

  \draw[fill=black] (0,3) circle (1.5pt)
  edge [<-] (-2,3);
  \draw[fill=black] (0,1) circle (1.5pt)
  edge [<-] (-2,1)
  edge [<-] node[above] {$x_2$} (-2,3);
  \draw[fill=black] (0,-1) circle (1.5pt)
  edge [<-] (-2,-1)
  edge [<-] node[above] {$x_2$} (-2,1);
  \draw[fill=black] (0,-3) circle (1.5pt)
  edge [<-] (-2,-3)
  edge [<-] node[above] {$x_2$} (-2,-1);

  \draw[fill=black] (2,3) circle (1.5pt)
  edge [<-] (0,3);
  \draw[fill=black] (2,1) circle (1.5pt)
  edge [<-] (0,1)
  edge [<-] node[above] {$x_3$} (0,3);
  \draw[fill=black] (2,-1) circle (1.5pt)
  edge [<-] (0,-1)
  edge [<-] node[above] {$x_3$} (0,1);
  \draw[fill=black] (2,-3) circle (1.5pt)
  edge [<-] (0,-3)
  edge [<-] node[above] {$x_3$} (0,-1);

  \draw[fill=black] (4,3) circle (1.5pt)
  edge [<-] (2,3);
  \draw[fill=black] (4,1) circle (1.5pt)
  edge [<-] (2,1)
  edge [<-] node[above] {$x_4$} (2,3);
  \draw[fill=black] (4,-1) circle (1.5pt)
  edge [<-] (2,-1)
  edge [<-] node[above] {$x_4$} (2,1);
  \draw[fill=black] (4,-3) circle (1.5pt)
  edge [<-] (2,-3)
  edge [<-] node[above] {$x_4$} (2,-1);

  \draw[fill=black] (6,3) circle (1.5pt)
  edge [<-] (4,3);
  \draw[fill=black] (6,1) circle (1.5pt)
  edge [<-] (4,1)
  edge [<-] node[above] {$x_5$} (4,3);
  \draw[fill=black] (6,-1) circle (1.5pt)
  edge [<-] (4,-1)
  edge [<-] node[above] {$x_5$} (4,1);
  \draw[fill=black] (6,-3) circle (1.5pt)
  edge [<-] (4,-3)
  edge [<-] node[above] {$x_5$} (4,-1);

  \node[draw,circle] (t) at (8,0) {$t$}
  edge [<-] (6,-3)
  edge [<-,color=gray] node[above] {$0$} (6,3)
  edge [<-,color=gray] node[above] {$0$} (6,1)
  edge [<-,color=gray] node[above] {$0$} (6,-1);
\end{tikzpicture}
\end{center} \caption{A \emph{commutative} ROABP for $\ESym^3_5$ (unlabelled edges have the label $1$).}
\label{fig:comm-for-esym}
\end{figure}
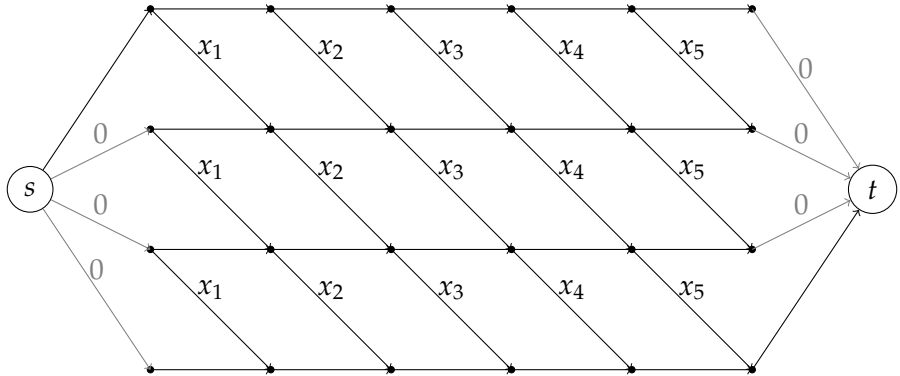

\begin{definition}[Elementary Symmetric Polynomials]
\label{defn:esym}
The $n$-variate elementary symmetric polynomial of degree $d$, denoted by $\ESym^d_{n} $ is defined as follows.
\begin{equation}\label{eq:ESym-definition}
  \ESym^d_{n}(\vecx) := \sum_{\substack{S \subset [n]\\ \abs{S} = d}} \prod_{i \in S} x_{i} \qedhere
\end{equation} 
\end{definition}
Following is a folklore construction (with a minor tweak) of an ROABP for $\ESym^d_n $ which is provably tight owing to the characterisation result by Nisan~\cite{N91} (see \autoref{sec:basics}).
We illustrate the construction for $n=5$ and $ d=3$ in the \autoref{fig:comm-for-esym} 
and give the general recipe here without a proof of correctness.


\begin{construction}\label{cons:comm-for-ESym}
For any $n,d \in \N$ such that $d \leq n$, we have the following.
\[
  \ESym^d_n(\vecx) = \inparen{M(x_1) M(x_2) \cdots M(x_n)}[1,d+1],
\]
where for all $i$, $M(x_i)$ is a $(d+1)\times(d+1)$ matrix such that $M(x_i)[k,k] = 1$ for all $1 \leq k \leq (d+1)$, and $M(x_i)[k,k+1] = 1$ for all $1 \leq k \leq d$; all other entries of $M(x_i)$ are zero.
\end{construction}
The matrix $M(x_i)$ can also be written as $(I + A x_i) $, where $A$ is a matrix with $1$s on its super-diagonal and zeros everywhere else, and $I$ is the identity matrix.
This gives the expression: $\ESym^d_n(\vecx) = \inparen{(I+A x_1) (I+A x_2) \cdots (I+A x_n)}_{(1,d+1)} = \trans{\vecu} \inparen{\prod_{i \in [n]} (I+A x_i)} \vecv $, for the obvious choice of $\vecu,\vecv \in \C^{(d+1)} $.

\noindent We can now make the following sequence of simple observations about this construction.
\begin{itemize}
\item All the coefficient matrices of the above ROABP: $I$ and $A$, commute with each other.
Thus, it is a commutative ROABP.
\item $(I+A x_1) (I+A x_2) \cdots (I+A x_n) = \sum_{0 \leq j \leq n} \ESym^j_n A^j = \sum_{0 \leq j \leq d} \ESym^j_n A^j $, since $A^j = 0 $ for all $j \geq (d+1)$.
\item For every $0 \leq j \leq d$, only the $j$th power of $A$ that has a $1$ in the $(1,1+j)$th entry. Therefore, the $(1,d+1)$th entry of $(I+A x_1) (I+A x_2) \cdots (I+A x_n)$ exactly computes the coefficient of $A^d $, which is $\ESym^d_n $.
\end{itemize}

This perspective along with elementary interpolation (\autoref{lem:interpolation}), then leads us to the following \emph{depth-3-multilinear} circuit for $\ESym^d_n $ of \emph{top fan-in} $(n+1)$ for all values of $d$, that is attributed to Ben-Or (\cite{SW01}).
This also happens to give the following nearly-optimal construction for a diagonal ROABP computing $\ESym^d_{n}$.
\begin{construction}\label{cons:diagonal-for-ESym}
For any $n,d \in \N$ and distinct $a_0,a_1,\ldots,a_n \in \C$, there exist constants $\beta_0,\beta_1,\ldots,\beta_n \in \C$ such that
\[
  \ESym^d_n(\vecx) = \sum_{0 \leq j \leq n} \beta_j (1 + a_j x_1) (1 + a_j x_2) \cdots (1 + a_j x_n) \qedhere
\]
\end{construction}

Just as the commutative ROABP for $\ESym^d_n(\vecx)$ leads us to Ben-or's construction of a diagonal ROABP, we also observe that the commutative ROABP computing $d$th power of an $n$-variate linear form $(x_1 + x_2 + \cdots + x_n)^d $ gives us the \emph{duality trick} of Saxena~\cite{S08b} (see e.g. \cite[Lemma 17.13]{S15}). We shall work with $(x_1 + x_2 + \cdots + x_n)^d $ for simplicity; all the ideas easily generalise to $d$th powers of arbitrary linear forms.



\begin{example}[Powers of linear forms]
\label{ex:pow-lin}
The $d$th powers of $n$-variate linear form is the polynomial $(x_1 + x_2 + \cdots + x_n)^d$. 
\end{example}

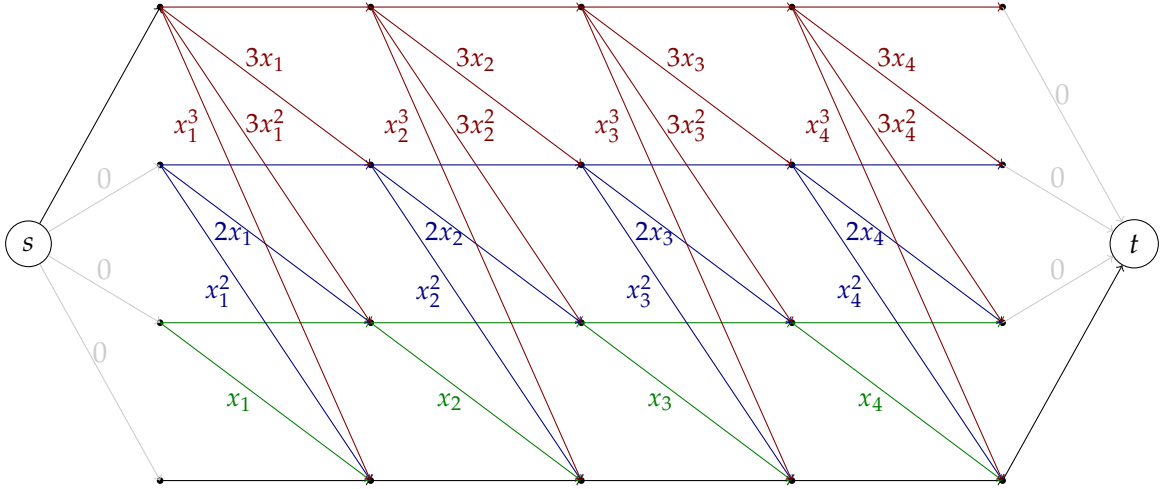
\begin{figure}[H]
\begin{center}
\begin{tikzpicture}[scale=0.7]

  \node[draw, circle] (s) at (-10.5,0) {$s$};

  \draw[fill=black] (-8,4.5) circle (1.5pt)
  edge [<-] (s);
  \draw[fill=black] (-8,1.5) circle (1.5pt)
  edge [<-, color=black!20!white] node[above] {$0$} (s);
  \draw[fill=black] (-8,-1.5) circle (1.5pt)
  edge [<-, color=black!20!white] node[above] {$0$} (s);
  \draw[fill=black] (-8,-4.5) circle (1.5pt)
  edge [<-, color=black!20!white] node[above] {$0$} (s);

  \draw[fill=black] (-4,4.5) circle (1.5pt)
  edge [<-, color=red!50!black] (-8,4.5);
  \draw[fill=black] (-4,1.5) circle (1.5pt)
  edge [<-, color=blue!50!black] (-8,1.5)
  edge [<-, color=red!50!black] node[pos=0.5, yshift=10] {{\small $3 x_1$}} (-8,4.5);
  \draw[fill=black] (-4,-1.5) circle (1.5pt)
  edge [<-, color=green!50!black] (-8,-1.5)
  edge [<-, color=blue!50!black] node[pos=0.65, yshift=-5] {{\small $2 x_1$}} (-8,1.5)
  edge [<-, color=red!50!black] node[pos=0.5, yshift=15] {{\small $3 x_1^2 $}} (-8,4.5);
  \draw[fill=black] (-4,-4.5) circle (1.5pt)
  edge [<-] (-8,-4.5)
  edge [<-, color=green!50!black] node[pos=0.5, xshift=-10] {{\small $x_1$}} (-8,-1.5)
  edge [<-, color=blue!50!black] node[pos=0.6, xshift=-10] {{\small $x_1^2 $}} (-8,1.5)
  edge [<-, color=red!50!black] node[pos=0.75, xshift=-10] {{\small $x_1^3 $}} (-8,4.5);

  \draw[fill=black] (0,4.5) circle (1.5pt)
  edge [<-, color=red!50!black] (-4,4.5);
  \draw[fill=black] (0,1.5) circle (1.5pt)
  edge [<-, color=blue!50!black] (-4,1.5)
  edge [<-, color=red!50!black] node[pos=0.5, yshift=10] {{\small $3 x_2$}} (-4,4.5);
  \draw[fill=black] (0,-1.5) circle (1.5pt)
  edge [<-, color=green!50!black] (-4,-1.5)
  edge [<-, color=blue!50!black] node[pos=0.65, yshift=-5] {{\small $2 x_2$}} (-4,1.5)
  edge [<-, color=red!50!black] node[pos=0.5, yshift=15] {{\small $3 x_2^2 $}} (-4,4.5);
  \draw[fill=black] (0,-4.5) circle (1.5pt)
  edge [<-] (-4,-4.5)
  edge [<-, color=green!50!black] node[pos=0.5, xshift=-10] {{\small $x_2$}} (-4,-1.5)
  edge [<-, color=blue!50!black] node[pos=0.6, xshift=-10] {{\small $x_2^2 $}} (-4,1.5)
  edge [<-, color=red!50!black] node[pos=0.75, xshift=-10] {{\small $x_2^3 $}} (-4,4.5);

  \draw[fill=black] (4,4.5) circle (1.5pt)
  edge [<-, color=red!50!black] (0,4.5);
  \draw[fill=black] (4,1.5) circle (1.5pt)
  edge [<-, color=blue!50!black] (0,1.5)
  edge [<-, color=red!50!black] node[pos=0.5, yshift=10] {{\small $3 x_3$}} (0,4.5);
  \draw[fill=black] (4,-1.5) circle (1.5pt)
  edge [<-, color=green!50!black] (0,-1.5)
  edge [<-, color=blue!50!black] node[pos=0.65, yshift=-5] {{\small $2 x_3$}} (0,1.5)
  edge [<-, color=red!50!black] node[pos=0.5, yshift=15] {{\small $3 x_3^2 $}} (0,4.5);
  \draw[fill=black] (4,-4.5) circle (1.5pt)
  edge [<-] (0,-4.5)
  edge [<-, color=green!50!black] node[pos=0.5, xshift=-10] {{\small $x_3$}} (0,-1.5)
  edge [<-, color=blue!50!black] node[pos=0.6, xshift=-10] {{\small $x_3^2 $}} (0,1.5)
  edge [<-, color=red!50!black] node[pos=0.75, xshift=-10] {{\small $x_3^3 $}} (0,4.5);

  \draw[fill=black] (8,4.5) circle (1.5pt)
  edge [<-, color=red!50!black] (4,4.5);
  \draw[fill=black] (8,1.5) circle (1.5pt)
  edge [<-, color=blue!50!black] (4,1.5)
  edge [<-, color=red!50!black] node[pos=0.5, yshift=10] {{\small $3 x_4$}} (4,4.5);
  \draw[fill=black] (8,-1.5) circle (1.5pt)
  edge [<-, color=green!50!black] (4,-1.5)
  edge [<-, color=blue!50!black] node[pos=0.65, yshift=-5] {{\small $2 x_4$}} (4,1.5)
  edge [<-, color=red!50!black] node[pos=0.5, yshift=15] {{\small $3 x_4^2 $}} (4,4.5);
  \draw[fill=black] (8,-4.5) circle (1.5pt)
  edge [<-] (4,-4.5)
  edge [<-, color=green!50!black] node[pos=0.5, xshift=-10] {{\small $x_4$}} (4,-1.5)
  edge [<-, color=blue!50!black] node[pos=0.6, xshift=-10] {{\small $x_4^2 $}} (4,1.5)
  edge [<-, color=red!50!black] node[pos=0.75, xshift=-10] {{\small $x_4^3 $}} (4,4.5);

  \node[draw, circle] (t) at (10.5,0) {$t$}
  edge [<-] (8,-4.5)
  edge [<-,color=black!20!white] node[above] {$0$} (8,-1.5)
  edge [<-,color=black!20!white] node[above] {$0$} (8,1.5)
  edge [<-,color=black!20!white] node[above] {$0$} (8,4.5);

\end{tikzpicture}
\end{center}
\caption{A \emph{commutative} ROABP for $(x_1+\cdots+x_4)^3$ (unlabelled edges have the label $1$).}
\label{fig:comm-for-power-of-linear}
\end{figure}

Consider the ROABP computing $(x_1 + x_2 + x_3 + x_4)^2 $ in \autoref{fig:comm-for-power-of-linear}; it will be convenient to index the vertices in each layer starting from zero.
We ensure that $j$th vertex in the $i$th layer, say $v_{i,j}$, has the property that the polynomial computed between $v_{i,j} $ and the \emph{sink vertex} $t$, is exactly $(x_i + \cdots + x_n)^{d-j} $.

The general construction can be described as follows.
\begin{construction}\label{cons:comm-for-power-of-linear}
For any $n,d \in \N$, we have the following.
\[
  (x_1 + x_2 + \cdots + x_n)^d = \inparen{M(x_1) M(x_2) \cdots M(x_n)}[1,d+1],
\]
where for each $i \in [n]$, $M(x_i)$ is a $(d+1) \times (d+1)$ matrix satisfying the following.
For all $0 \leq k \leq d$ and all $0 \leq \ell \leq (d-k)$, $M(x_i)[k,k+\ell] = \binom{d-k}{\ell} x_i^{\ell} $.
\end{construction}

Just as in \autoref{cons:comm-for-ESym}, we now write the matrix $M(x_i)$ as a univariate over $x_i $ with matrix coefficients.
\[
  M(x_i) = I + A^1 x_i + \frac{A^2}{2!} x_i^2 + \frac{A^3}{3!} x_i^3 + \cdots + \frac{A^d}{d!} x_i^d,
\]
where $A$ is a $(d+1) \times (d+1)$ such that for all $0 \leq i \leq d$, $A[i,i+1] = (d-i)$, with all other entries set to zero.\\

We now make the following simple observations.
\begin{itemize}\itemsep 0pt
\item All the \emph{coefficient matrices} of the above ROABP: $I$ and powers of $A$, commute with each other. Thus, \autoref{cons:comm-for-power-of-linear} is also a \emph{commutative ROABP}.
\item For every $0 \leq j \leq d$, only the $j$th power of $A$ that has a $1$ in the $(1,1+j)$th entry. Therefore, the $(1,d+1)$th entry of $M(x_1) M(x_2) \cdots M(x_n)$ computes the coefficient of $A^d $ divided by $d!$. This can be seen as follows.
\end{itemize}

\newcommand\scalemath[2]{\scalebox{#1}{\mbox{\ensuremath{\displaystyle #2}}}}

\[
  M(x_i) =\scalemath{0.8}{\begin{bmatrix}
    \binom{d}{0} & 0 & 0 & 0 & \cdots & 0\\
    0 & \binom{d-1}{0} & 0 & 0 & \cdots & 0\\
    \vdots & \vdots & \vdots & \vdots & \ddots & \vdots\\
    0 & 0 & 0 & 0 & \cdots & 0\\
    0 & 0 & 0 & 0 & \cdots & 0\\
    0 & 0 & 0 & 0 & \cdots & \binom{0}{0}\\
  \end{bmatrix}}
  x_i^0 +
  \scalemath{0.8}{\begin{bmatrix}
    0 & \binom{d}{1} & 0 & 0 & \cdots & 0\\
    0 & 0 & \binom{d-1}{1} & 0 & \cdots & 0\\
    \vdots & \vdots & \vdots & \vdots & \ddots & \vdots\\
    0 & 0 & 0 & 0 & \cdots & 0\\
    0 & 0 & 0 & 0 & \cdots & \binom{1}{1}\\
    0 & 0 & 0 & 0 & \cdots & 0\\
  \end{bmatrix}} x_i^1 + \cdots +
 \scalemath{0.8}{\begin{bmatrix}
    0 & 0 & 0 & 0 & \cdots & \binom{d}{d}\\
    0 & 0 & 0 & 0 & \cdots & 0\\
    \vdots & \vdots & \vdots & \vdots & \ddots & \vdots\\
    0 & 0 & 0 & 0 & \cdots & 0\\
    0 & 0 & 0 & 0 & \cdots & 0\\
    0 & 0 & 0 & 0 & \cdots & 0\\
  \end{bmatrix}}
  x_i^d.
\]

%
\[
  \therefore M(x_i) = \cdots + \frac{1}{j!}
  \scalemath{0.8}{\begin{bmatrix}
    0 & d & 0 & 0 & \cdots & 0\\
    0 & 0 & d-1 & 0 & \cdots & 0\\
    \vdots & \vdots & \vdots & \vdots & \ddots & \vdots\\
    0 & 0 & 0 & 0 & \cdots & 0\\
    0 & 0 & 0 & 0 & \cdots & 1\\
    0 & 0 & 0 & 0 & \cdots & 0\\
  \end{bmatrix}}
  ^j x_i^j + \cdots = I + A^1 x_i + \frac{A^2}{2!} x_i^2 + \frac{A^3}{3!} x_i^3 + \cdots + \frac{A^d}{d!} x_i^d.
\]

Rather surprisingly, applying interpolation now leads us to the following sum-of-products-of-univariates for $(\sum_i x_i)^d $, that exactly matches ``the duality trick'' shown by Saxena~\cite{S08b}!
\begin{construction}\label{cons:diagonal-for-power-of-linear}
For any $n,d \in \N$ and distinct $a_{0},a_1,\ldots,a_{nd} \in \C$, there exist $\beta_{0},\beta_1,\ldots,\beta_{nd} $ for which the following holds.
\[
  (x_1 + x_2 + \ldots + x_n)^d = \sum_{0 \leq j \leq nd} \beta_j \prod_{i \in [n]} \inparen{1 + a_j x_i + \frac{a_j^2}{2!} x_i^2 + \frac{a_j^3}{3!} x_i^3 + \cdots + \frac{a_j^d}{d!} x_i^d} \qedhere
\]
\end{construction}

As the coefficient matrices of diagonal ROABPs are diagonal matrices it is not difficult to observe that they are exactly \emph{sums-of-products-of-univariates}.
Thus, from \autoref{ex:pow-lin} (i.e. duality trick), we observe that diagonal ROABPs can efficiently simulate \emph{diagonal depth $3$ circuits} (a.k.a. \emph{depth-$3$ powering circuits}) denoted by $\SES$.
That is, $\SES(n,d,s) \subseteq \diagROABP(n,d,O(n,d,s))$.
Also, a separation between these two classes is known due to the exponential lower bound from \cite{NW97} for $x_1\ldots x_n$ against the model $\SES$.
In essence, we have the following containments between classes\footnote{We have more intricate relationships between classes concerning ROABPs. See \autoref{subsec:roabp-landscape}},
where each $\calc$ stands for the class of $n$-variate, degree-$d$ polynomials whose $\calc$-size is $\poly(n,d)$.
\[
  \SES \subsetneq \diagROABP \subseteq \commROABP \subseteq  \ROABP[\forall] \subsetneq \ROABP[\exists]
\]

Looking at the above hierarchy, we firstly realise that nearly optimal separations are known at the two ``extremes'', but nothing is known about the intermediate levels.
Further, since the intermediate levels are far more algebraically structured (coefficient matrices arising from special commutative algebras), it is reasonable to expect finer separations for these classes.
Unfortunately, all the lower bounds that we know for diagonal and commutative ROABPs are those that are known for $\ROABP[\forall]$.

Secondly, even though diagonal ROABPs (\emph{sum-of-products-of-univariates}) may be of independent interest as they subsume $\SES$ circuits, they are also interesting from the point of view of polynomial identity testing.
Owing to the algebraic structure of their coefficients, one can expect efficient PIT algorithms for these classes.
But again, the best PIT algorithms that we know for diagonal and commutative ROABPs are those we know for $\ROABP[\forall]$.
We discuss more about polynomial identity testing algorithms for these classes in \autoref{sec:pit}.

\subsection{Our Results}

We now move to the central questions addressed in this article.
In particular, we wish to understand if the classes $\commROABP$ and $\diagROABP$ are equal up to polynomial factors; this can be more formally stated as follows.
\begin{question}\label{quest:comm-diag-ROABP}
Given an $n$-variate, individual degree $d$ polynomial $f(\vecx)$ computable by a width $w$ commutative ROABP(i.e. $f\in \commROABP(n,d,w)$), does there exist a diagonal ROABP computing $f$ of width $\poly(n,d,w)$?
\end{question}


A measure that is often used to prove lower bounds against structured models (e.g. almost every lower bound against $\SES$, and more recently \cite{LST21}) is the \emph{dimension of partial derivatives}, a complexity measure which was introduced by Nisan and Wigderson~\cite{NW97} (see \autoref{defn:dimension-of-partials}).
For any polynomial $f \in \mathbb{C}[\vecx]$, the partial derivative complexity of $f$ (denoted by $\partials(f)$) is the dimension of the space spanned by \emph{all} the partial derivatives of $f$.
Nisan and Wigderson~\cite{NW97} observed that any $n$-variate, degree $d$ polynomial $f(\vecx)$ that has a $\SES$ circuit of size $s$ has $\partials(f) \leq s(d+1)$.
Therefore it is natural to ask whether the $\SES$-size of every polynomial $f$ is polynomially related to its dimension of partial derivatives.
We formalize this question as follows.
\begin{question}\label{quest:partials-and-waring}
Does there exist a \emph{constant} $c$ such that for any $n$-variate, degree-$d$ polynomial $f(\vecx)$ with $\partials(f) \leq s$, we have that the smallest $\SES$ circuit that computes $f(\vecx)$ has size at most $(nds)^c$?
\end{question}

The size of the smallest $\SES$ circuit for a polynomial is a well studied notion called the \emph{Waring rank of $f$} (denoted by $\waring(f)$).
\autoref{quest:partials-and-waring} essentially asks if the Waring rank and the dimension partial derivatives of a polynomial are same up to polynomial factors.
Unfortunately, at the moment we do not know the answers to either \autoref{quest:comm-diag-ROABP} or \autoref{quest:partials-and-waring}.
However, our main result gives a rather surprising connection between \autoref{quest:comm-diag-ROABP} and \autoref{quest:partials-and-waring}.
Specifically, we show that an positive answer to \autoref{quest:partials-and-waring} answers \autoref{quest:comm-diag-ROABP} in the affirmative!

\begin{restatable}{theorem}{CommROABPToDiagROABP}
\label{thm:comm-roabp-to-diag-roabp}
For any $n,r \in \N$, let $S(r,m)$ denote the smallest $\SES$-size required to compute any $r$-variate polynomial $f$ with $\partials(f) \leq m$.\\
Then for all $n,d,w \in \N$, $\commROABP(n,d,w) \subseteq \diagROABP\inparen{n,d,S(w^2,w^2)nw^4}$.
\end{restatable}
\begin{remark}
It can be inferred from our proof that a super-polynomial separation between the models of $\commROABP$ and $\diagROABP$ will yield an explicit polynomial that witnesses a super-polynomial separation between dimension of partial derivatives and Waring rank.
We elaborate on this in \autoref{rem:connection}.
\end{remark}

A different (and perhaps equally surprising) consequence of \autoref{thm:comm-roabp-to-diag-roabp} is that a super-polynomial separation between commutative ROABPs and diagonal ROABP will also give a similar separation between dimension of partial derivatives and Waring rank.
Note that not only do we not know the answers to \autoref{quest:comm-diag-ROABP} or \autoref{quest:partials-and-waring}, it is somewhat frustrating that we do not even know of a candidate polynomial that could potentially separate these classes.
We expect that our analysis of these models that goes into proving the result above could help in making some progress in either of these questions.

\subsection{An overview of the proof}
\label{subsec:overview}
We start by asking when diagonal ROABPs can efficiently simulate commutative ROABPs.
This question naturally leads us to study properties of matrices that commute with each other.
In particular, we analyse \emph{commutative rings} generated by matrices that commute with each other.

\noindent \textbf{A very high level overview.}
The results of Marinari, {\Moller}, Mora~\cite{MMM93}, and {\Moller} and Stetter~\cite{MS95} provide a characterisation of commutative rings of $w \times w$ matrices in terms of polynomials whose \emph{dimension of partial derivatives} is at most $\poly(w)$.
In the special case when these matrices are all diagonal, the same polynomials happen to have \emph{Waring rank} at most $w$.
Further, we observe that if the polynomials corresponding to a $n$-variate, width-$w$ commutative ROABP have \emph{Waring rank} at most $s$, then it can be simulated by a diagonal ROABP of width $\poly(n,w,s)$.
This is essentially our main result.
We now explain the characterisation given by \cite{MMM93} and \cite{MS95} in a bit more detail.

\paragraph*{Characterising rings of matrices}
Consider the ring generated by a $w \times w$ matrix $A$, given by $\C[A] := \set{ q(A) : q(t) \in \C[t]}$.
The ring has at most $w$ \emph{linearly independent} matrices, as the characteristic polynomial of $A$ gives a way to express $A^w $ as a linear combination of lower powers of $A$.
In fact, the ring $\C[A]$ is characterised by the \emph{ideal} of all polynomials that are divisible by the \emph{minimal polynomial of $A$} (see \autoref{fact:quotient-ring-minimal-polynomial}).
This characterisation has an appropriate analogue for general matrix rings, as follows.

Suppose that $A_1,\ldots,A_r \in \C^{w \times w} $ commute with each other, and let $\C[A_1,\ldots,A_r]$, defined as $\set{ g(A_1,\ldots,A_r) : g(\vect) \in \C[\vect] } $, be the ring generated by them\footnote{Any ring of $w \times w$ matrices is generated by at most $w^2 $ matrices.}.
Analogous to the univariate (singly-generated) case, we then consider the \emph{ideal of dependencies} for the matrices $A_1,\ldots,A_r $: $J = \set{ p(\vect) \in \C[\vect] : p(A_1,\ldots,A_r)=0 }$.
As it turns out, $\C[A_1,\ldots,A_r]$ is indeed characterised by the ideal $J$ (see \autoref{lem:quotient-ring-ideal-dependencies}).

Before delving further into the ideal of dependencies, we remark a structural property of polynomials that admit a diagonal ROABP of a certain width.

\noindent \textbf{Understanding diagonal ROABPs.}
Consider the diagonal ROABP (depth-3 multilinear circuit) for the \emph{elementary symmetric polynomial} $\ESym_{n,d} $ that is attributed to Ben-Or (see e.g. \cite{SW01}).
One first constructs the polynomial $g(t,\vecx) := (1 + t x_1)(1 + t x_2)\cdots(1 + t x_n)$, and then obtains $\ESym_{n,d} $ as the coefficient of $t^d $ in $g(t,\vecx)$, using interpolation.
It turns out that any diagonal ROABP computing a polynomial $f(\vecx)$ can similarly be seen as expressing $f$ as a linear combination of evaluations of a \emph{low-degree} $g(t,\vecx)$ that is a ``product of univariates'' (see \autoref{obs:diagonal-roabp-alt-view}).
Here, the number of evaluations needed is \emph{equal} to the width of the ROABP.
Moreover the converse of this statement is also true, thus giving us an equivalent formulation for diagonal ROABPs.

Therefore, we analyse the ideal $J$ with the goal of expressing the corresponding commutative ROABP as a \emph{sum of $\vect$-evaluations} of some $G(\vect,\vecx) = G_1(\vect,x_1) \cdot G_2(\vect,x_2) \cdots G_n(\vect,x_n) $.

\noindent \textbf{The ideal of dependencies.}
Let us first make our statement about $\C[A_1,\ldots,A_r]$ being characterised by $J$ a bit more precise: there is a \emph{ring-isomorphism} between $\C[A_1,\ldots,A_r]$ and the \emph{quotient ring} $\sfrac{\C[\vect]}{J}$.
Therefore it is crucial to understand $J$ (and $\sfrac{\C[\vect]}{J}$) to understand the ring of matrices, in order to move towards the above mentioned goal.

Let $p(t)$ be the minimal polynomial of some matrix $A$, and consider the ideal $\ideal{p}$.
If $p(t) = (t - 5)^3 $, then we know that any $q(t)$ belongs to $\ideal{p}$ \emph{if and only if} the first 3 derivatives of $q(t)$ vanish at $t = 5$; i.e. $q(5) = q'(5) = q''(5) = 0$.
In general, for $p(t) = (t - a_1)^{e_1} (t - a_2)^{e_2} \cdots (t - a_k)^{e_k}$, membership in the ideal $\ideal{p}$ is \emph{characterised} by the first $e_i $ derivatives vanishing at $t = a_i$, for \emph{each} $i = 1,2,\ldots,k$.
Moreover, the polynomial ``$q(t) \bmod p(t)$'' can be obtained by applying a \emph{linear transformation} on the evaluations of the $e_1,\ldots,e_k $ derivatives at the respective points $a_1,\ldots,a_k $.

We now extend this understanding to the multivariate setting.
We already have the correct analogue for $\ideal{p}$, which we call the ideal of dependencies $J$.
Next, we need a characterisation for ``$g(\vect) \bmod J$'' in terms of some derivatives of $g(\vect)$ evaluated at some points related to $J$.
While these choices were quite clear in the univariate setting from $p$; the multivariate setting requires a little more care.
Fortunately for us, the works of Marinari, M\"{o}ller, Mora~\cite{MMM93}, and M\"{o}ller and Stetter~\cite{MS95} provide an adequate solution.

Firstly, observe that $J$ has a finite \emph{variety} (common zeroes of all polynomials in $J$).
Thus the variety $\V(J)$ is a good multivariate analogue for the set of evaluation points.
The other ingredient that we require is a compatible notion of ``multiplicity of $J$'' at a point $\bar{\alpha}$ in its variety.
For this, \cite{MMM93} look at the set of all \emph{partial derivative operators} (see \autoref{defn:derivative-operators}) which map \emph{every} polynomial in $J$ to a polynomial that vanishes at $\bar{\alpha}$.
These operators form a vector space over $\C$, and the ``multiplicity of $J$ at $\bar{\alpha}$'' is then defined as the \emph{dimension} of this vector space.

In the univariate setting, the multiplicity of $q$ at a point $a_i$ is defined as the \emph{highest} number $e_i $ such that the first $e_i $ derivatives of $q$ vanish at the point $a_i $.
Thus, one can naturally identify a ``highest derivative'', with the other derivatives being its ``down-shifted versions''.
Analogously, the derivative operator space corresponding to $J$ and a point $\vecv \in \V(J)$ is \emph{closed under taking down-shifts} (see \autoref{defn:down-closed-spaces}).
An ideal $J$ with $\V(J) = \set{\bar{\alpha}_1,\ldots,\bar{\alpha}_k}$, is then captured by a collection of $z$ vector spaces of derivative operators $\Delta_1, \Delta_2, \ldots, \Delta_k $, in the following sense (see \autoref{lem:MMM-ideal-derivative-correspondence-weak}).
\begin{itemize} \itemsep 0pt
  \item For each $i \in [k]$, $\Delta_i $ corresponds to the point $\bar{\alpha}_i $ and is down-closed.
  \item Dimension of the quotient ring $\sfrac{\C[\vect]}{J}$ is $w = \dim(\Delta_1) + \dim(\Delta_2) + \cdots + \dim(\Delta_k)$.
  \item Let $\set{D_{i,1},\ldots,D_{i,w_i}}$ be a basis of $\Delta_i $. Then there exists a map $\Phi : \C^{w} \rightarrow \sfrac{\C[\vect]}{J} $ such that for any polynomial $q(\vect)$, $\Phi$ maps the $w$ values: $\set{D_{i,j}(q)(\vecv_i)}$, to the ``remainder polynomial'' $\inparen{q(\vect) \bmod I}$.
\end{itemize}
Further, {\Moller} and Stetter~\cite{MS95} show that the map $\Phi$ stated above is just a linear transformation (see \autoref{lem:dual-basis-ideals}).

\noindent \textbf{Consequences for ROABPs.}
We now outline the proof of our main result (\autoref{thm:comm-roabp-to-diag-roabp}).
\begin{itemize}
  
  \item Given a commutative ROABP $f(\vecx) = \trans{\vecb} \cdot \prod_{i \in [n]} \inparen{A_{i,0} + A_{i,1} x_i + \cdots + A_{i,d} x_i^{d} } \cdot \vecc $ of width $w$, we define $F(\vecx) := \prod_{i \in [n]} \inparen{A_{i,0} + A_{i,1} x_i + \cdots + A_{i,d} x_i^{d} }$ to be a matrix of polynomials.
  Then, $f(\vecx)$ is just a linear combination (given by $\vecb \trans{\vecc}$) of the entries of $F$.

  \item We then identify a set of matrices $A_1,\ldots,A_r $ that generate the coefficient-matrix-ring; i.e. $\C[A_1,\ldots,A_r] = \C[A_{1,0},\ldots,A_{1,d},\ldots,A_{n,d}]$.
  As we can always use the coefficient matrices themselves, and because we are dealing with $w \times w$ matrices, $r \leq \min(w^2,n(d+1))$.

  \item Let $J$ be the ideal of dependencies for $A_1,\ldots,A_r $ and suppose the normal set of $J$ (see \autoref{defn:normal-set}) has size, say $m \leq w^2 $.
  Then each $A_{i,j} $ is a polynomial in $A_1,\ldots,A_r $ that has $\leq m$ monomials.

  \item For each $i,j$, suppose $G_{i,j}(t_1,\ldots,t_r)$ is the polynomial such that $G_{i,j}(A_1,\ldots,A_r) = A_{i,j} $; the entries of $A_{i,j} $ are linear combinations of $\vect$-coefficients of $G_{i,j} = (G_{i,j} \bmod J) $.
  Then we observe that $G(\vect,\vecx) := \prod_{i \in [n]} \inparen{G_{i,0}(\vect) + G_{i,1}(\vect) x_i + \cdots + G_{i,d}(\vect) x_i^{d} }$, with $G(A_1,\ldots,A_r,\vecx) = F(\vecx)$.
  This means that even $f(\vecx)$ is a linear combination of the $\vect$-coefficients of $(G(\vect,\vecx) \bmod J)$, since it is a linear combination of the entries of $F(\vecx)$.
  We prove this in \autoref{lem:comm-roabp-ideal-form}.

  \item Now let $\V(J) = \set{\vecv_1,\ldots,\vecv_k}$ and for each $\ell \in [k]$ let $\set{D_{\ell,1},\ldots,D_{\ell,m_{\ell}}}$ be a basis for the derivative operator space corresponding to $\vecv_{\ell} $.
  Then from the results of \cite{MMM93,MS95} we get that for any $g(\vect)$, every $\vect$-coefficient of $(g(\vect) \bmod J)$ is a fixed linear combination of the $m$ values given by $(D_{\ell,\ast}(g))(v_{\ell}) $.

  \item This brings us one step away from our goal of expressing $f(\vecx)$ as a linear combination of $\vect$-evaluations of some $G(\vect,\vecx)$ which is a product of univariates.
  What we need is a way to express each of $(D_{\ell,\ast}(G))(\vecv)$ as a linear combination of $\vect$-evaluations of $G(\vect,\vecx)$.

  \item It turns out that the number of evaluations of $G(\vect,\vecx)$ required to compute $(D_{\ell,\ast}(G))(\vecv)$ is $\poly(\deg(h_{\ell,\ast}),\waring(h_{\ell,\ast}))$, where $h_{\ell,\ast}$ is the \emph{polynomial corresponding to $D_{\ell,\ast} $} (see paragraph below \autoref{defn:derivative-operators}).
  This is a non-trivial fact; we prove it in \autoref{lem:functional-waring-rank}.

  \item Finally, since each space $\Delta_{\ell} $ is \emph{down-closed}, we have that the dimension of partial derivatives $\partials(h_{\ell,\ast}) \leq \dim(\Delta_{\ell}) \leq m $ for each $h_{\ell,\ast} $.
  Therefore, using the hypothesis that $\waring(h) = \poly(r,\partials(h))$ for any $r$-variate $h$, we get that $(D_{\ell,\ast}(G))(\vecv)$ can be expressed as a linear combination of $\poly(r,\partials(h_{\ell,\ast}),\deg(h_{\ell,\ast})) = \poly(r,m)$ evaluations of $G(\vect,\vecx)$ for each $D_{\ell,\ast} $.
  
  \item Combining all the above observations, we can see that the hypothesis implies that $f(\vecx)$ can indeed be written as a linear combination of $\poly(r,m) = \poly(n,d,w)$ evaluations of $G(\vect,\vecx)$, thereby proving \autoref{thm:comm-roabp-to-diag-roabp}.
\end{itemize}

\subsection{Landscape of ROABP classes}
\label{subsec:roabp-landscape}

As mentioned earlier, although \autoref{thm:comm-roabp-to-diag-roabp} relates \autoref{quest:comm-diag-ROABP} and \autoref{quest:partials-and-waring}, the answer to both these questions remain unknown.
In this regard, we would like to conjecture that the answer to both questions is false.  
\begin{conjecture}
\label{conj:comm-and-diag}
There exists an explicit $n$-variate, degree-$d$ polynomial $f(\vecx)$ with a commutative ROABP of width $\poly(n,d)$, such that any diagonal ROABP computing $f$ requires width $n^{\omega(1)}$.
\end{conjecture}

\begin{conjecture}
\label{conj:partials-and-waring}
There exists an explicit $n$-variate polynomial $f(\vecx)$ of degree $d = \poly(n)$ such that $\partials(f)=\poly(n)$ but $\waring(f) = n^{\omega(1)}$. 
\end{conjecture}

Even though many would agree that \autoref{conj:comm-and-diag} and \autoref{conj:partials-and-waring} are probably true, we do not even know of any candidate polynomial that will witness the truth of this conjecture. 
In relation to this, we remark that the following statement can be inferred from our proof of \autoref{thm:comm-roabp-to-diag-roabp}.
If there exists a commutative ROABP of width $\poly(n,d)$ computing an $n$-variate, degree-$d$ polynomial $f$, which requires diagonal ROABPs of super-polynomial width, then the commutative ROABP for $f$ gives a different explicit polynomial $h$ that has polynomial dimension of partial derivatives, but has super-polynomial Waring rank (see \autoref{rem:connection} for details).
As a result, even a candidate polynomial for proving \autoref{conj:comm-and-diag} remains unknown.

In the context of \autoref{conj:comm-and-diag}, we note the following connection between diagonal ROABPs and tensor rank.
\begin{remark}
\label{rem:diag-ROABP}
Observe that the width of a diagonal ROABP exactly captures the \emph{tensor rank} of the corresponding \emph{tensor}.
A tensor $T : [d]^{n}  \rightarrow \mathbb{C}$ of order\footnote{Commonly used term in the literature about tensors; not be confused with the order of an ROABP.} $n$ can naturally be viewed as a polynomial $f_T = \sum_{\veci \in [d]^n} T(i_1,\ldots, i_n) x_1^{i_1}\cdots x_n^{i_n}$.
The (tensor) rank of any $T$ (denoted by $\TR(T)$) is the smallest $r$ such that $T$ can be expressed as sum of $r$ elementary tensors.
Thus for any tensor $T$, $\TR(T)=r$ if and only if $f_T(\vecx)$ can be expressed as sum of $r$ many products of univariates; which immediately implies $\diagROABP(n,d,w) = \{ f_T \in \mathbb{C}[\vecx] \mid \TR(T) \leq w \}$.
Obtaining strong lower bounds on the rank of explicit tensors is a major open problem in algebraic complexity theory, where the goal is to obtain an explicit tensor $T$ of order-$n$ such that $\TR(T)= d^{n(1-o(1))}$ (see e.g. \cite{Raz10}).
\end{remark}
\autoref{rem:diag-ROABP} tells us that proving strong width lower bounds against diagonal ROABPs could potentially imply lower bounds on the rank of explicit tensors.
While this could partially explain why there are no separations between diagonal ROABPs and commutative or ``all-order'' ROABPs, it is also worth mentioning that order-$n$ tensors for a \emph{growing parameter} $n$ are rarely studied in the context of tensor rank lower bounds.

\medskip

With regard to \autoref{conj:partials-and-waring}, we briefly discuss some known results about the problem of computing the dimension of the partial derivative space.

Shitov~\cite{S16} showed that given any degree $3$ polynomial $f$ in its sparse representation, computing $\waring(f)$ is $\NP$-hard, by reducing it to computing the tensor rank of order $3$ \emph{symmetric tensors}.
On the other hand, when a polynomial $f$ is presented in its sparse representation (as sum of monomials), Garc{\'{\i}}a{-}Marco, Koiran, Pecatte and Thomass\'{e}~\cite{GKPT17} prove that computing the dimension of the partial derivative space is $\#\P$-hard (not known to be $\#\P$-complete).
Thus, even though computing Waring rank is a hard problem, it is not quite clear if disproving \autoref{conj:partials-and-waring} goes against it.
Moreover, it is possible that Waring rank is easy to \emph{approximate} up to polynomial factors, which is all that a disproof of \autoref{conj:partials-and-waring} would imply.
On a related note, Kayal~\cite{K12} gave a randomised $\poly(n,d)$-time algorithm to compute the waring rank of an $n$-variate, degree-$d$ polynomial that is given as a blackbox (in the \emph{non-degenerate case}).

\medskip

\begin{figure}[t]
\begin{center}
\begin{tikzpicture}[transform shape,scale=0.8]
 \node [draw, rectangle, thick, align=center] (roabp-exists) at (-3.5,14) {$\poly$-width ROABP in some order\\
 $\ROABP[\exists](n,d,\poly(n,d))$ \\
 (\autoref{defn:ROABP-in-some-order})};
 \node [draw, rectangle, thick, align=center] (roabp-all) at (-3.5,10.5) {$\poly$-width ROABP in every order  \\
$\ROABP[\forall](n,d,\poly(n,d)) $ \\
 (\autoref{defn:ROABP-in-any-order})};
 \node [draw, rectangle, thick, align=center] (comm-roabp) at (-1.5,7) {$\poly$-width commutative ROABP\\ 
$\mathsf{comm\mbox{-}ROABP}(n,d,\poly(n,d))$ \\
  (\autoref{defn:commutative-ROABP})};
  \node [draw, rectangle, thick, align=center] (diag-roabp) at (4,3) {$\poly$-width diagonal ROABP \\  
  $\mathsf{diag\mbox{-}ROABP}(n,d,\poly(n,d)) $ \\

  (\autoref{defn:diagonal-ROABP})};
  \node [draw, rectangle, thick, align=center] (waring) at (-2,-1.2) {$\poly$-size $\SES$ circuits \\ 
  $\{ f\in \mathbb{C}^{\leq d}[\vecx] \mid \waring(f)\leq \poly(n,d) \}$ \\ 
  
  (\autoref{defn:depth-3-powering}) };
  \node [draw, rectangle, thick, align=center] (partials) at (-7,4.4) {$\poly$ dimension of partial derivates \\ 
  $\{f\in \mathbb{C}^{\leq d}[\vecx] \mid \partials(f)\leq  \poly(n,d)\ \}$ \\ 
  
  (\autoref{defn:dimension-of-partials}) };
  \node [draw, rectangle, thick, align=center] (border-diagonal) at (5,6.7) {$\poly$-width border of diagonal ROABP \\  $\overline{\mathsf{diag\mbox{-}ROABP}}(n,d,\poly(n,d))$ \\
  (\autoref{defn:border-diag})
  }; 
  \draw [->] (roabp-all) -- node [draw, rectangle, fill=white] {$\neq$} (roabp-exists);
  \draw [->] (comm-roabp) -- node [draw, rectangle, fill=white] {$\stackrel{?}{=}$} (roabp-all);
  \draw [->] (diag-roabp) -- node (diag-comm) [draw, rectangle, fill=white] {$\stackrel{?}{=}$} (comm-roabp);
  \draw [->] (waring) -- node [draw, rectangle, fill=white] {$\neq$} (diag-roabp);
  \draw [->] (waring) -- node (waring-partials) [draw, rectangle, fill=white] {$\stackrel{?}{=}$} (partials);
  \draw [->] (partials) -- node [draw, rectangle, fill=white] {$\neq$} (roabp-all);
  \draw [->] (diag-roabp) -- node [draw, rectangle, fill=white] {$\stackrel{?}{=}$} (border-diagonal);
  \draw [->] (border-diagonal) -- node [draw, rectangle, fill=white] {$\stackrel{?}{=}$} (roabp-all);
  \draw [<-, double, color=red, thick] (diag-comm) -- (waring-partials) ;
\draw [<-, double, color=red, thick] (diag-comm) -- node [draw, circle, fill=white] {$\Rightarrow$} (waring-partials) ;
\end{tikzpicture}
\end{center}
\caption{The ROABP landscape: edges denote bottom-up inclusion, \autoref{thm:comm-roabp-to-diag-roabp} is in red.} 
\label{fig:roabp-landscape}
\end{figure}
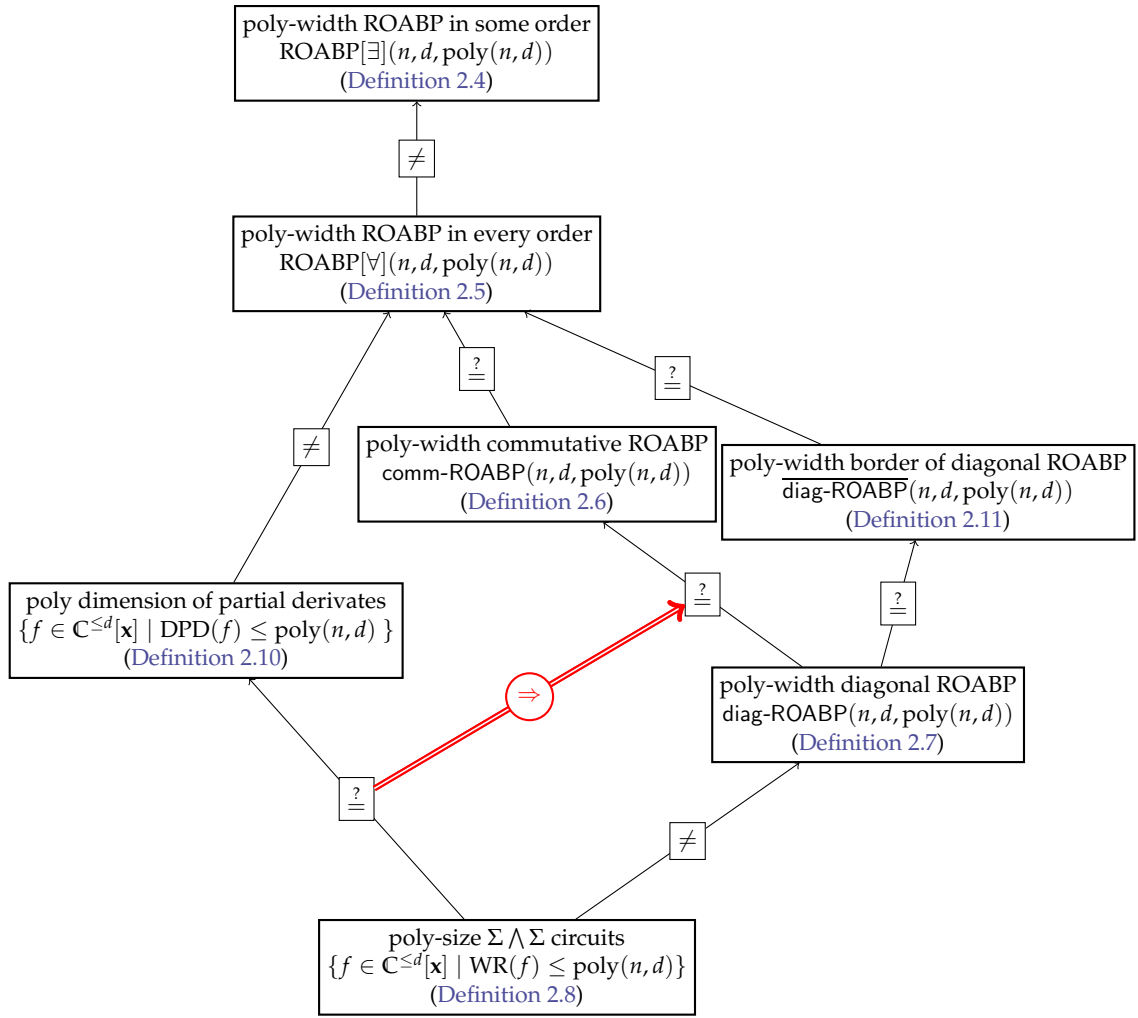

Although the results in this article entirely concern \autoref{quest:partials-and-waring} and \autoref{quest:comm-diag-ROABP}, there are several other interesting open questions surrounding the landscape of complexity classes involving ROABPs.
We discuss these interconnections between ROABP classes now, and later illustrate them in \autoref{fig:roabp-landscape}.

Let us consider the class of polynomials computed by ROABPs that remain unchanged by interchanging layers in the branching program\footnote{The class $\ROABP[\forall](n,d,w)$ has been studied in the context of PIT, and is sometimes called \emph{commutative ROABPs} in some works (e.g. \cite{GKS17}). We use a different notation to avoid any ambiguity.}.
We prefer to use the term \emph{layer-commutative ROABPs} (denoted by $\mathsf{layer}\mbox{-}\commROABP(n,d,w)$) to denote the  class of $n$-variate degree $d$ polynomials computed by an ROABPs such that if $f = u^T M_1(x_1)\cdots M_n(x_n)v$ then the matrices of univariate polynomials $M_1,\ldots, M_n$ commute.
That is, $M_i(x_i)M_j(x_j) = M_j(x_j)M_i(x_i)$ for all $i,j\in [n]$.
We can immediately see that $\mathsf{layer}\mbox{-}\commROABP(n,d,w) \subseteq \ROABP[\forall](n,d,w)$, and further $\commROABP(n,d,w) \subseteq \mathsf{layer}\mbox{-}\commROABP(n,d,w)$.
This then leads us to the following two open questions whose answer seems unclear at the moment.
\begin{question}
\label{quest:comm-and-forall}
\begin{enumerate}\itemsep 0pt
\item Are $\mathsf{layer}\mbox{-}\commROABP(n,d,w)$ and $\ROABP[\forall](n,d,w)$ equivalent up to a polynomial blow-up in the width $w$?
\item Are $\commROABP(n,d,w)$ and $\mathsf{layer}\mbox{-}\commROABP(n,d,w)$  equivalent up to a polynomial blow-up in the width $w$?
\end{enumerate}
\end{question}
We hope that a better understanding the algebra associated with commutative ROABPs may shed light on the answers to above questions.

\medskip
Along with the complexity of computing polynomials exactly, another notion that is considered in algebraic complexity theory and more specifically in \emph{geometric complexity theory}, is \emph{border complexity} of polynomials.
Let $\mathcal{C}$ be a class of polynomials. We say that $f$ is in the class $\overline{C}$ (border of $\mathcal{C}$), if $f$ can be ``arbitrarily-approximated'' by a circuit in $\mathcal{C}$.
That is, there exists a polynomial $g(\epsilon) \in \mathbb{C}(\epsilon)$ in class $\mathcal{C}$ such that $f = \lim_{\epsilon \to 0} g$.
The \emph{border-complexity} of $f$ is then at most the size of the circuit computing $g$.
Clearly, $\mathcal{C} \subseteq \overline{\mathcal{C}}$. Understanding whether $\mathcal{C}= \overline{\mathcal{C}}$ for interesting classes such as $\VP$ and $\VBP$ are major open problems in algebraic complexity theory.
Here, we are interested in the case when $\mathcal{C} = \diagROABP(n,d,w)$ (defined in \autoref{defn:border-diag}).
\begin{question}
\label{quest:diag-and-bdiag}
Does the model of diagonal ROABPs require super-polynomial width to simulate the border class $\overline{\diagROABP}$?
\end{question}
As $\diagROABP(n,d,w) = \{ f_T \in \mathbb{C}[\vecx] \mid  \mathsf{TR}(T) \leq w \}$, we have $\overline{\diagROABP(n,d,w)} = \{ f_T \in \mathbb{C}[\vecx] \mid  \underline{\mathsf{TR}}_{\mathbb{C}}(f) \leq w \}$. Here, $\underline{\mathsf{TR}}(f)$ denotes the border rank of tensors.
Border rank of tensors is studied extensively in several contexts for instance, border rank of \emph{matrix multiplication tensor} is used to obtain bounds on the arithmetic complexity of matrix multiplication.
In this setting, the order of the tensor is usually bounded by a constant, and this setting slightly deviates from the main theme algebraic circuit complexity.

\medskip

It can be checked that just like $\commROABP(n,d,w)$, $\overline{\diagROABP}(n,d,w)$ is also contained in $\ROABP[\forall](n,d,w)$ (because $\ROABP$-complexity is characterised by rank, which is a continuous measure).
However, it is unclear if these two ways of ``generalising'' diagonal ROABPs have different computational powers.
This brings us to the following question.
\begin{question}
\label{quest:comm-and-bdiag}
Are the classes $\overline{\diagROABP}(n,d,w)$ and $ \commROABP(n,d,w)$ equivalent up to polynomial factors?
\end{question}

Note that \autoref{quest:comm-and-bdiag} is linked to the question of understanding $\commROABP(n,d,w)$ and $\ROABP[\forall](n,d,w)$ in \autoref{quest:comm-and-forall}.
Also, answering this question in the affirmative is similar in spirit to the recent ``de-bordering'' results due to Dutta et al.~\cite{DDS21}.
They proved that the border of constant \emph{top fan-in} depth three circuits is contained in the class $\VBP$.
Here, \autoref{quest:comm-and-bdiag} is essentially asking if for the class of diagonal ROABPs (albeit with unbounded fan-in), the border is contained in a much simpler class of commutative ROABPs?
However, answering this in the negative could potentially be as hard as (or even harder than) separating commutative ROABPs from diagonal ROABPs.
In fact, it is not even clear if these two classes should be comparable (contained in one another).
We believe that any answer to \autoref{quest:comm-and-bdiag} would be an interesting development in algebraic complexity theory.

\medskip

We summarize all the models and the interconnections between the structured ROABP classes in \autoref{fig:roabp-landscape}.

\section{Preliminaries}
\label{sec:preliminaries}

In this section, we formally define the classes of polynomials and other algebraic models of computation that we study in this paper.
We work over the field of complex numbers $\C$, unless mentioned otherwise.
We begin with some basic notation and the definitions of \emph{algebraic circuits} and \emph{algebraic branching programs}.

\paragraph*{Notation}
\label{para:notation}
\begin{itemize}
  \item We use the shorthand $[n]$ to denote the set $\set{1,2,\ldots,n}$.
  \item We use boldface letters like $\vecx,\vect,\vecA,\vecb$, to denote sets (and vectors). The individual elements (and coordinates) are denoted by indexed versions of the same characters: $\vecA = \set{A_1,\ldots,A_r}$.\\
  Whenever the size of these sets is not clear from context, we denote them using subscripts: $\vecx_{[n]} = \set{x_1,\ldots,x_n} $.
  \item For a polynomial $f(\vecx)$, we denote \emph{support of $f$} the set of monomials appearing in $f$ with a nonzero coefficient by $\supp(f)$.
  \item For $\vecx = \set{x_1,\ldots,x_n}$, and any vector $\vece \in \N^{n} $, we use the shorthand $\vecx^{\vece} $ to denote the monomial $x_1^{e_1}x_2^{e_2}\cdots x_n^{e_n} $.
  \item For a polynomial $f(\vecx)$ and a monomial $\vecx^{\vece} $, we use $\partial_{\vece}f $ to denote the partial derivative $\frac{\partial^{\abs{\vece}} f}{\partial x_1^{e_1} \cdots \partial x_n^{e_n}}$.
  \item For a matrix $M$, $M[i,j]$ denotes its $(i,j)$th entry.
\end{itemize}

\begin{definition}[Algebraic circuits] \label{defn:alg-circuits}
  An \emph{algebraic circuit} is specified by a directed acyclic graph, with leaves (in-degree zero; also called \emph{inputs}) labelled by field constants or variables, and internal nodes labelled by $+$ or $\times$.
  The nodes with out-degree zero are called the \emph{outputs} of the circuit.
  Computation proceeds in the natural way, where inductively each $+$ gate computes the sum of its children and each $\times$ gate computes the product of its children.

  The \emph{size} of the circuit is defined as the number of nodes in the underlying graph.
\end{definition}

\begin{definition}[Algebraic Branching Programs]
\label{defn:abp}
An algebraic branching program is a layered, directed graph. There are two special vertices, source $s$ and sink $t$ which are the only  vertices in the first and last layers, respectively. All the edges in the graph are from one layer to the consecutive layer. Each edge is labelled by a univariate polynomial in the underlying variables over the underlying field. Each path from $s$ to $t$ computes the product of the edge labels and the ABP computes the sum of all the paths from $s$ to $t$. Then, any ABP can be viewed as a product of matrices (each matrix having univariate polynomials as its entries) and the ABP computes the $(1, 1)$th entry of the matrix product. The maximum number of vertices in a single layer (dimension of the largest matrix in the product) is called its \emph{width}. The size of the ABP is the total number of vertices in it.
\end{definition}

We now define the various structured ROABPs and other related classes that are the main objects of interest in our paper.

We start by defining the basic model of ROABPs.
\begin{definition}[Read-once Oblivious ABPs]
\label{defn:roabp}
Over the field $\C$ of complex numbers, a \emph{read-once oblivious algebraic branching program} or an \emph{ROABP}, computes an $n$-variate, \emph{individual} degree $d$ polynomial using a matrix-vector product of the following form.
\[
   R(\vecx) = \trans{\vecu} \cdot M_1(x_{\sigma(1)}) \cdot M_2(x_{\sigma(2)}) \cdots M_n(x_{\sigma(n)}) \cdot \vecv
\]
where
\begin{itemize}\itemsep 0pt
  \item For each $i \in [n]$, the matrix $M_i(x_{\sigma(i)})$ has entries that are univariates of degree $\leq d$ in the variable $x_{\sigma(i)} $,
  \item $\vecu \in \C^{w_0} $, $M_1(x_{\sigma(1)}) \in (\C[x_{\sigma(1)}])^{w_0 \times w_1} $, \ldots, $M_i(x_{\sigma(i)}) \in (\C[x_{\sigma(i)}])^{w_i \times w_{i+1}} $, \ldots, $\vecv \in \C^{w_n} $,
  \item the \emph{width} $w$ of the ROABP $R$ is defined as $w = \max\set{w_0,w_1,\ldots,w_n}$,
  \item the permutation $\sigma$ is called as \emph{the order of the ROABP $R$}. \qedhere
\end{itemize}
\end{definition}

The following two subclasses of polynomials then follow naturally from the definition of ROABPs.
\begin{definition}[ROABPs in some order]\label{defn:ROABP-in-some-order}
For $n,d,w \in \N$, an $n$-variate polynomial $f(\vecx)$ of \emph{individual} degree $d$ is said to have an ROABP of width $w$ \emph{in the order $\sigma \in S_n$}, if there exists a width $w$ ROABP $R(\vecx) $ that computes $f(\vecx)$ in the order $\sigma$.
We denote the class of such polynomials by $\ROABP[\sigma](n,d,w)$.\\
Further, we use $\ROABP[\exists](n,d,w)$ to denote the class of polynomials that have a width $w$ ROABP in \emph{some} order. That is, $\ROABP[\exists](n,d,w) = \bigcup_{\sigma \in S_n} \ROABP[\sigma](n,d,w)$.
\end{definition}

We can then extend this definition naturally as follows.
\begin{definition}[ROABPs in every order]\label{defn:ROABP-in-any-order}
For $n,d,w \in \N$, an $n$-variate polynomial $f(\vecx)$ of \emph{individual} degree $d$ is said to have an ROABP of width $w$ \emph{in every order}, if for all permutations $\sigma \in S_{n}$, there exists a width $w$ ROABP $R_{(\sigma)}(\vecx) $ that computes $f(\vecx)$ in the order $\sigma$.\\
We denote this class of polynomials by $\ROABP[\forall](n,d,w)$.
\end{definition}

Now, based on the properties of the \emph{coefficient matrices}, we define the two subclasses of ROABPs that \autoref{thm:comm-roabp-to-diag-roabp} talks about.


\begin{definition}[Commutative ROABPs]\label{defn:commutative-ROABP}
An $n$-variate, individual degree $d$ ROABP of width $w$ is called a \emph{commutative ROABP} if its coefficient matrices are all $w \times w$ matrices that are (pairwise) commutative.\\
We refer of the class of polynomials computed by such ROABPs by $\commROABP(n,d,w)$.
\end{definition}
\begin{definition} [Diagonal ROABPs] \label{defn:diagonal-ROABP}
An $n$-variate, individual degree $d$ ROABP of width $w$ is called a \emph{diagonal ROABP} if its coefficient matrices are $w \times w$ \emph{diagonal} matrices. We refer of the class of polynomials computed by such ROABPs by $\diagROABP(n,d,w)$.
\end{definition}

Further, we define other concepts about polynomials like \emph{depth-$3$ powering circuits}, \emph{Waring rank} and \emph{Tensor rank}, since we talk about the connections between them and subclasses of ROABPs defined above.
\begin{definition}[Depth 3 powering circuits ($\Sigma\wedge\Sigma$)]
\label{defn:depth-3-powering}
A depth 3 powering circuit of size $s$, computes an $n$-variate, (total) degree $d$ polynomial as an $\C$-linear combination of $s$ terms, each of which is a $\leq d$th power of an $\C$-linear form in the underlying variables $x_1,\ldots,x_n $.\\
That is, vectors $\veca_1,\ldots,\veca_s \in \C^{n+1} $, constants $\beta_1,\ldots,\beta_s $, and $d_1,d_2,\ldots,d_s \in \set{0,\ldots,d} $, define the following $n$-variate, degree-$d$, size $s$ \emph{depth $3$ powering circuit}.
\[
   C(\vecx) = \sum_{i \in [s]} \beta_i \inparen{a_0 + a_1 x_1 + a_2 x_2 + \cdots + a_n x_n}^{d_i}  \qedhere
\]
\end{definition}

\begin{definition}[Waring rank]
\label{defn:waring-rank}
For an $n$-variate, degree-$d$ polynomial $f(\vecx) \in \C[\vecx]$, the \emph{Waring rank of $f$} is defined to be the size of the smallest depth $3$ powering circuit that computes it. We will denote the Waring rank of a polynomial $f$ by $\waring(f)$.
\end{definition}

\begin{definition}[Dimension of partial derivatives]
\label{defn:dimension-of-partials}
For an $n$-variate polynomial $f(\vecx) \in \C[\vecx]$, its \emph{dimension of partial derivatives} or $\partials(f)$, is defined as ${\partials(f) = \dim\inparen{\Fspan_{\C}\set{ \partial_{\vece}f : \vece \in \N^n}}}$.
Here, $\partial_{\vece}f $ denotes the partial derivative $\frac{\partial^{\abs{\vece}} f}{\partial x_1^{e_1} \cdots \partial x_n^{e_n}}$.
\end{definition}

Finally, we define the border of diagonal ROABPs as follows, which coincides with the definition of commonly known definition of \emph{border-tensor-rank}.
\begin{definition}[Border of diagonal ROABPs]
\label{defn:border-diag}
For any polynomial $f(\vecx) \in \mathbb{C}[\vecx]$, $f(\vecx)$ is in the class $\overline{\diagROABP(n,d,w)}$ if there exists a polynomial $g \in \mathbb{C}(\epsilon)$ in the class $\diagROABP(n,d,w)$ such that $f = \lim\limits_{\epsilon\to 0} g$. 
\end{definition}

\section{Algebraic structure of commutative ROABPs}
\label{sec:algebra-background}

This section is aimed at equipping the reader with the algebraic-geometric concepts about \emph{rings generated by commuting matrices}, that are required to understand the results in \cite{MMM93} and \cite{MS95} (\autoref{lem:MMM-ideal-derivative-correspondence-weak} and \autoref{lem:dual-basis-ideals}).
It is therefore largely expository, and readers who are comfortable with these concepts may skip it.


We start by analysing rings generated by a single matrix in \autoref{subsec:univariate-rings}, and then extend our observations to general rings of matrices in \autoref{subsec:multivariate-rings}.

\subsection{Rings generated by a single matrix}
\label{subsec:univariate-rings}

For any matrix $A \in \C^{w \times w} $, the commutative ring generated by $A$ that is denoted by $\C[A]$, is the set of all matrices that can be written as univariate polynomials in terms of $A$.
In other words, $\C[A] := \set{ p(A) : p(t) \in \C[t]}$.

Observe that the matrices $I(=A^0), A, A^2, \ldots, A^w $ satisfy the linear dependency that is given by the \emph{characteristic polynomial of $A$}: $\det(A - tI) \in \C[t]$.
Thus, $\C[A]$ is a vector space (over $\C$) of dimension at most $w$.

In fact the dimension of $\C[A]$ could be even smaller, and it is captured by the degree of the \emph{minimal polynomial of $A$}: the smallest degree polynomial $p(t)$ such that $p(A)$ is the zero matrix; and the ideal generated by $p$, $\ideal{p} := \set{ q(t) \in \C[t] : q(t)\text{ is divisible by }p(t) }$, characterises the ring $\C[A]$.
The following fact formalises this relationship.

\begin{fact}
\label{fact:quotient-ring-minimal-polynomial}
Let $A \in \C^{w \times w} $ and let $p(t) \in \C[t]$ be its minimal polynomial.
Then the ring generated by $A$, $\C[A]$, is \emph{isomorphic} to the quotient ring $\sfrac{\C[t]}{\ideal{p}}$.
\end{fact}
\begin{proof}
Define $\Phi : \C[t] \rightarrow \C[A] $ such that $\Phi(q(t)) = q(A)$ for any $q$.
Then the following facts together show that the restriction of $\Phi$ on $\sfrac{\C[t]}{\ideal{p}}$ is a ring isomorphism by the \emph{first ring isomorphism theorem} (see e.g. \cite{DF99}).
\begin{itemize}\itemsep 0pt
  \item $\Phi$ is a ring homomorphism: $\Phi(q_1 + q_2 \cdot q_3) = (q_1 + q_2 \cdot q_3)(A) = q_1(A) + q_2(A) \cdot q_3(A) $.
  \item $\Phi$ is \emph{onto}: Trivially follows from the definition of $\C[A]$.
  \item $\ker \Phi = \ideal{p}$: Suppose $\Phi(q) = 0$. Then $q(A) = 0$, which implies that $q(t) = p(t) \cdot q'(t)$ as $p(t)$ is the minimal polynomial of $A$. \qedhere
\end{itemize}
\end{proof}

Let us now focus on the quotient ring of the ideal generated by an arbitrary polynomial $p(t)$; we shall later rephrase our findings in terms of matrices.

Suppose $p(t) = (t-\alpha_1)^{e_1} (t-\alpha_2)^{e_2} \cdots (t-\alpha_z)^{e_z}$, of degree $m = \sum_u e_u $.
Since we are working over $\C$, this is true without loss of generality.
Let $p_u$ be the polynomial $(t - \alpha_u)^{e_u} $, for each $u \in [z]$.
Then any polynomial $q(t)$ is divisible by $p_u $ whenever $\alpha_{u}$ is a root of $q(t)$ and its first $(e_u - 1)$ derivatives.
In fact, $q(t)$ is divisible by $p = \prod_u p_u $, exactly when the above condition holds for each $u \in [z]$.

\begin{fact}
\label{fact:derivatives-univariate-ideals}
A polynomial $q(t)$ is divisible by $p(t) = \prod_{u \in [z]} (t - \alpha_u)^{e_u} $ if and only if:
\[
  \forall u \in [z], \qquad q(\alpha_u) = \frac{\partial q}{\partial t}(\alpha_u) = \frac{\partial^2 q}{\partial t^2}(\alpha_u) = \cdots = \frac{\partial^{{e_u} - 1} q}{\partial t^{{e_u} - 1}}(\alpha_u) = 0.
\]
\end{fact}
In other words, the $\sum_u e_u = m $ values obtained by evaluating the appropriate derivatives of $q$ at the corresponding roots of $p$, tell us whether $p$ divides $q$.
These evaluations of derivatives in fact give us some more information about $q$ with respect to the ideal $\ideal{p}$, which we now see.

\noindent \textbf{Derivatives characterise the quotient ring.}
For any polynomials $p(t), q(t)$ we define the ``remainder polynomial'' $q(t) \bmod p(t)$ as follows.
\[
    q(t) \bmod p(t) = \tilde{q}(t) \text{, such that } q(t) = q'(t)p(t) + \tilde{q}(t), \text{ with } \deg(\tilde{q}) < \deg(p)
\]
Suppose $p(t)$ is a polynomial of degree $m$, then $\tilde{q}(t)$ is clearly a polynomial of degree at most $m-1$.
It turns out that the $d$ evaluations of derivatives of $q$ given in \autoref{fact:derivatives-univariate-ideals} completely determine $\tilde{q}$.

\begin{fact}
\label{fact:derivatives-characterise-quotients}
Suppose $p(t) = \prod_{u \in [z]} (t - \alpha_u)^{e_u} $ has degree $m$, then there exist $m^2$ constants $\set{\gamma^{a}_{u,v}} \subset \C$ such that for any polynomial $q(t)$, we have
\[
    \forall 0 \leq a \leq m-1, \qquad \tilde{q}_a = \sum_{\substack{u \in [z]\\v \in [e_u]}} \gamma^{(a)}_{u,v} \cdot \frac{\partial^{v} q}{\partial t^{v}}(\alpha_u),
\]
where $\tilde{q}(t) := \sum_{0 \leq j \leq m-1} \tilde{q}_j t^j = q(t) \bmod p(t)$.
\end{fact}

\subsection{General commutative matrix rings}
\label{subsec:multivariate-rings}

The above observations about ``univariate'' rings can be summarised as follows. Firstly, any matrix ring is isomorphic to the quotient ring of an ideal, where this ideal contains all polynomial dependencies that the generator matrix satisfies (\autoref{fact:quotient-ring-minimal-polynomial}); thus every matrix in the ring corresponds to a polynomial modulo this ideal.\\
Secondly, the remainder of any polynomial $q$ with respect to this ideal is completely determined by the evaluations of certain derivatives of $q$ at appropriate points (\autoref{fact:derivatives-characterise-quotients}).

We shall now see the multivariate analogues of the above facts, which tell us about rings generated by multiple commuting matrices.

To fix some notation, suppose that we have been given the $w \times w $ matrices $A_1,\ldots,A_r $ that all commute with each other.
These matrices therefore generate a commutative ring of matrices denoted by $\C[A_1,\ldots,A_r]$, whose algebraic properties we shall now provide.

\subsubsection{Matrix rings as quotient rings of ideals}

Recall that for the ring $\C[A]$, the corresponding ideal was $\ideal{p(t)}$, where $p$ was the minimal polynomial of $A$.
The ideal $\ideal{p(t)}$ precisely contains all the polynomials $q(t)$ for which $q(A) = 0$.
Therefore a natural choice for the multivariate ideal is the \emph{ideal of dependencies of $A_1,\ldots,A_r$}, $J := \set{ q(t_1,\ldots,t_r) \in \C[\vect] : q(A_1,\ldots,A_r) = 0 }$.
Indeed, the quotient ring of $J$ is isomorphic to $\C[A_1,\ldots,A_r]$.

\begin{lemma}
\label{lem:quotient-ring-ideal-dependencies}
Suppose $A_1,A_2,\ldots,A_r \in \C^{w \times w} $ are mutually commutative, and let $J$ be their \emph{ideal of dependencies} inside the $r$-variate polynomial ring $\C[\vect]$.
Then $\C[A_1,\ldots,A_r] $ is isomorphic to $\sfrac{\C[\vect]}{J} $.
\end{lemma}
\begin{proof}
Similar to the proof of \autoref{fact:quotient-ring-minimal-polynomial}, we define the map $\Phi : \C[\vect] \rightarrow \C[A_1,\ldots,A_r] $, which maps $q(\vect)$ to the matrix $q(A_1,\ldots,A_r)$.
This naturally defines the (restricted) map $\phi : \sfrac{\C[\vect]}{J} \rightarrow \C[A_1,\ldots,A_r]$, with $\phi(\tilde{q}) = \tilde{q}(\vecA)$.

The following facts are now easy to verify for $\Phi$, which together prove that $\phi$ is an isomorphism by the \emph{first ring isomorphism theorem} (see e.g. \cite{DF99}).
\begin{itemize}\itemsep 0pt
  \item $\Phi$ is a ring homomorphism: $\Phi(q_1 + q_2 \cdot q_3)$ $ = (q_1 + q_2 \cdot q_3)(A_1,\ldots,A_r) $ $= q_1(\vecA) + q_2(\vecA) \cdot q_3(\vecA)$ $ = \Phi(q_1) + \Phi(q_2) \cdot \Phi(q_3) $.
  \item $\Phi$ is \emph{onto}: Trivially follows from the definition of $\C[\vecA]$.
  \item $\ker \Phi = J$: Suppose $\Phi(q) = 0$. Then $q(\vecA) = 0$, which implies that $q(\vect) \in J$. \qedhere
\end{itemize}
\end{proof}

We note an important property of the ideal $J$, before moving on to the next part.
Notice that the minimal polynomials of each of the matrices $A_1,\ldots,A_r $, say $p_1(t_1),p_2(t_2),\ldots,p_r(t_r) $ are elements of $J$.
This means that $J$ contains univariate polynomials in each of its underlying variables.
Thus, the set of common zeroes of polynomials in $J$, also known as the \emph{variety of $J$}(denoted by $\V(J)$), is finite.
One way to see this is that $\V(J) \subseteq \roots(p_1) \times \roots(p_2) \times \cdots \times \roots(p_r)$, where $\roots(p_i)$ denotes the constants in $\C$ where $p_i $ vanishes, and $\times$ denotes the \emph{Cartesian product} of sets.
Such ideals are called \emph{zero dimensional ideals}, because their variety is a zero dimensional set in the ambient space $\C^{r} $.

\begin{definition}[Zero-dimensional ideals]
\label{defn:zero-dimensional-ideals}
An ideal $J \subseteq \C[\vect]$ is called \emph{zero-dimensional} if its variety is finite; i.e. $\abs{\V(J)} < \infty$.
\end{definition}

\subsubsection{Quotient rings of zero dimensional ideals}

Since we are interested in zero dimensional ideals $J$, we shall now assume that $\V(J) = \set{v_1,\ldots,v_{z}}$ for some $z \in \N$.

Arguably, the statements we have discussed till this point are fairly well-known. 
But we believe that most of the ideas we shall now see are not as commonly known, especially in the theoretical computer science community.
We remark that much of the non-trivial ideas and proofs in this section (\autoref{sec:algebra-background}) belong to previous works~\cite{MMM93,MS95}.

Taking a cue from \autoref{fact:derivatives-characterise-quotients}, for a zero-dimensional ideal $J$ we expect the ``multiplicities'' of the points in its variety $\V(J)$ to help us find the correct derivatives.
In this case, the commonly used definition of multiplicity for multivariate polynomials: multiplicity of $w$ means \emph{all} partial derivatives of order $< w$ vanish, turns out to be a little too coarse.
In order to formally introduce the suitable definition, we need the following notion of \emph{derivative operators}, which are like polynomials whose monomials are partial derivatives.

\begin{definition}[Derivative operators]
\label{defn:derivative-operators}
A \emph{derivative operator} on $\C[t_1,\ldots,t_r]$ is a $\C$-linear combination of \emph{finitely many} partial derivatives of the form $\partial_{\veca} : \C[\vect] \rightarrow \C[\vect] $, where $\veca \in \N^r $.

The operator $D = \sum_{\veca} \gamma_{\veca} \partial_{\veca} $ naturally maps a polynomial $q(\vect) \in \C[\vect]$, to $\inparen{\sum_{\veca} \gamma_{\veca} \cdot \partial_{\veca}q(\vect)} $ which we denote by $D(q)$.
\end{definition}
Any polynomial $h(\vect)$ naturally defines a derivative operator $D_h := \sum_{\veca \in \supp(h)} \coeff_h(\veca) \partial_{\veca} $.
Likewise, one can talk about the polynomial that underlies a derivative operator.

In \autoref{fact:derivatives-characterise-quotients}, the set of derivative-evaluations that characterise the ideal generated by a $p = (t - \alpha)^e $, are evaluations at $\alpha$ of derivatives with respect to the monomials $\set{t^{e-1}, t^{e-2}, \ldots, t, 1}$; for multiple factors we take the union of the evaluations for each factor.
In particular, there is a ``maximum'' derivative $\sfrac{\partial^e}{\partial t^{e}}$, and the other derivatives are obtained by ``down-shifting'' it (similar to taking all possible derivatives of the underlying monomial).
This observation leads us to define the following notion of \emph{shifts} of derivatives and derivative operators.

\begin{definition}[Shifts of derivatives and derivative operators]
\label{defn:shifts-of-derivatives}
For a partial derivative $\partial_{\vece} : \C[\vect] \rightarrow \C[\vect] $ and a vector $\veca \geq \bar{0}$, we define the \emph{$\veca$-shift of $\partial_{\vece}$}, denoted by $\sigma_{\veca}(\partial_{\vece}) $, as follows.
\begin{equation*}
   \sigma_{\veca}(\partial_{\vece}) :=
   \begin{cases*}
      \frac{\vece !}{(\vece - \veca) !} \cdot \partial_{\vece - \veca} & if $\veca \leq \vece$,\\
      0                                                                & otherwise.
    \end{cases*}
\end{equation*}

\noindent The definition naturally extends to \emph{$\veca$-shift of $D_h$}, denoted by $\sigma_{\veca}(D_h) $, as follows.
\[
    \sigma_{\veca}(D_h) := \sum_{\vece : \vece \geq \veca} \coeff_{\vece}(h) \cdot \sigma_{\veca}\inparen{\partial_{\vece}} = \sum_{\vece : \vece \geq \veca} \coeff_{\vece}(h) \cdot \frac{\vece !}{(\vece - \veca) !} \cdot \partial_{\vece - \veca} \qedhere
\]
\end{definition}

\noindent The following observations about derivative operators and their shifts will be useful.
\begin{observation}
\label{obs:shift-to-derivative}
For any derivative operator $D_h $ and vector $\veca$, $\sigma_{\veca}(D_h) = D_{\partial_{\veca}(h)} $.
\end{observation}

\begin{observation}
\label{obs:product-rule-operators}
For any derivative operator $D_h $ and polynomials $p(\vect),q(\vect)$, we have the following.
\[
    D_h(p \cdot q) =  \sum_{\veca} \frac{1}{\veca !} \cdot \partial_{\veca}(p) \cdot \sigma_{\veca}(D_h)(q) = \sum_{\veca} \frac{1}{\veca !} \cdot \partial_{\veca}(p) \cdot D_{\partial_{\veca}(h)}(q)
\]
\end{observation}

\begin{proof}
\begin{align*}
  D_h(p \cdot q)        &= \sum_{\vece \in \supp(h)} \coeff_{\vece}(h) \partial_{\vece}(p \cdot q) \\
  \text{(Product rule)} &= \sum_{\vece \in \supp(h)} \coeff_{\vece}(h) \sum_{\veca : \veca \leq \vece} \binom{\vece}{\veca} \cdot \partial_{\veca}(p) \cdot \partial_{\vece - \veca}(q) \\
  \text{(Rearranging)}  &= \sum_{\veca} \partial_{\veca}(p) \cdot \frac{1}{\veca !} \cdot \inparen{ \sum_{\vece \in \supp(h) : \vece \geq \veca} \inparen{ \frac{\vece !}{(\vece - \veca)!} \cdot \coeff_{\vece}(h)} \partial_{\vece - \veca}(q)} \\
  \text{(\autoref{defn:shifts-of-derivatives})}  &= \sum_{\veca} \frac{1}{\veca !} \cdot \partial_{\veca}(p) \cdot \sigma_{\veca}(D_h)(q) \\
  \text{(\autoref{obs:shift-to-derivative})}  &= \sum_{\veca} \frac{1}{\veca !} \cdot \partial_{\veca}(p) \cdot D_{\partial_{\veca}(h)}(q) \qedhere
\end{align*}
\end{proof}

In the language of shifts of derivative operators, we can say that the set of derivatives with respect to $\set{t^e, t^{e-1}, \ldots, t, 1}$ is \emph{down-closed}: closed under taking shifts.
The following definitions then follow naturally.

\begin{definition}[Down-closed spaces of derivative operators]\label{defn:down-closed-spaces}
A $\C$-vector space of derivative operators $\Delta$ is said to be \emph{down-closed} if for all $D \in \Delta $, any shift $D'$ of $D$, also belongs to $\Delta$.
\end{definition}

\begin{definition}[Closure of an operator]\label{defn:closure-of-operator}
For a polynomial $h(t_1,\ldots,t_r) \in \C[\vect]$ and the corresponding derivative operator $D_h $, we define the \emph{closure of $D_h$} as follows.

$\Delta(h) := \set{ D_{\partial_{\vece}(h)} : \vece \in \N^{r}, \partial_{\vece}(h) \neq 0 } $.
\end{definition}

\noindent Ideals with a single point in their variety and closed spaces of derivative operators have the following interesting connection, similar to a univariate ideal $\ideal{(t - \alpha)^e}$.
\begin{lemma}
\label{lem:operator-spaces-of-ideals}
Let $J \in \C[t_1,\ldots,t_r] $ be an ideal with $\V(J) = \set{\bar{\alpha}}$, then the set $\Delta(J)$ of derivative operators defined by $\Delta(J):= \set{ D \in \C[\partial t_1,\ldots,\partial t_r] : \forall g \in J, D(g)(\bar{\alpha}) = 0 } $ a closed vector space.
\end{lemma}
\begin{proof}
Firstly, for all $D_1,D_2 $, and $\beta \in \C$, $ (\beta D_1 + D_2)(f)(\bar{\alpha})$ $ = \beta D_1(f)(\bar{\alpha}) + D_2(f)(\bar{\alpha}) = 0 $, just by linearity of differentiation.
So $\Delta(J)$ is a vector space over $\C$.

To see that it is closed, suppose $D_h \in \Delta(J) $ for a polynomial $h(\vect)$, and let $i \in [r]$ be such that the partial derivative $h' := \sfrac{\partial h}{\partial t_i} \neq 0$.
Then using \autoref{obs:product-rule-operators}, for any $g \in J$ we have that $ D_h(t_i \cdot g)(\bar{\alpha}) = (t_i \cdot D_{h}(g) + 1 \cdot D_{h'}(g))(\bar{\alpha}) = v_i \cdot D_{h}(g)(\bar{\alpha}) + 1 \cdot D_{h'}(g)(\bar{\alpha})$.
Now since $J$ is an ideal, $g \in J$ implies that $t_i \cdot g \in I$ and therefore $D_h(t_i \cdot g)(\bar{\alpha})=0$; and $D_{h}(g)(\bar{\alpha}) = 0$ because $g \in J$ and $D_h \in \Delta(J) $.
Thus, $D_{h'}(g)(\bar{\alpha}) = 0$ for any $D_{h} \in \Delta(J)$ and $i \in [r]$ such that $\sfrac{\partial h}{\partial t_i} \neq 0$.
The closure under an arbitrary shift $\veca$ then follows by induction on the $\veca$.
\end{proof}

\noindent We are now ready to state the following result which follows from the work of Marinari, {\Moller} and Mora~\cite[Theorem 2.6]{MMM93}, which is a suitable multivariate analogue for \autoref{fact:derivatives-univariate-ideals}.

\begin{lemma}[Zero dimensional ideals and derivative operator spaces]
\label{lem:MMM-ideal-derivative-correspondence-weak}
Suppose an ideal $J \subseteq \C[\vect] $ has variety $\V(J) = \set{\bar{\alpha}_1,\ldots,\bar{\alpha}_z}$ and $\dim_{\C}\inparen{\sfrac{\C[\vect]}{J}} = m$.
Then there exist closed spaces of derivative operators $\Delta_1,\ldots,\Delta_z $ of dimensions $m_1, \ldots,m_z $ with $\sum_u m_u = m $, such that for any polynomial $g(\vect) \in \C[\vect]$ we have that $g \in J$, if and only if $\forall u \in [z], \forall D \in \Delta_u: D(g)(\bar{\alpha}_u) = 0$.
\end{lemma}

Thus, every zero-dimensional ideal is characterised by a set of closed spaces of derivative operators, where the number of spaces is equal to the size of the variety.
Next, we see how one can obtain ``$g \bmod J$'' given the $\sum_u m_u = m$ derivative-evaluations corresponding to the $z$ bases of $\Delta_1,\ldots,\Delta_z $.
To that end, we first formalise what $g \bmod J$ means and then state a result from \cite{MS95} that provides the above solution.

\subsubsection{Matrices and polynomials in the quotient ring}

When dealing with univariate polynomials, it is quite straightforward to define $q(t) \bmod p(t)$ as $r(t) $, such that $q(t) = q'(t)p(t) + r(t)$ for some polynomial $q'(t)$ with $\deg(r) < \deg(p)$.
This is because we intuitively identify $r(t)$ to be ``less than'' $p(t)$ since it has smaller degree, and thus the concepts of division and remainders extend naturally.
However, things are a little more tricky for multivariate polynomials: e.g. which monomial is ``smaller''? $x^2 $ or $y^2 $?

We therefore need to fix a consistent way of comparing any two given monomials; we need a \emph{monomial ordering}: a total ordering on monomials that ``respects'' division/multiplication (see e.g. \cite[Chapter 2]{CLO07}).
We shall skip the formal definition of a monomial ordering, and just work with the ``dictionary ordering'' or \emph{lexicographic ordering}: $\vect^{\veca} \ltmon \vect^{\veca'} $ if the smallest $i \in [r]$ with $a_i \neq a'_i $ is such that $a_i < a'_i $.
Using the monomial ordering $\ltmon$, we can define the \emph{leading monomial} of a polynomial, and then \emph{leading monomials of $J$} for an ideal $J$.
\begin{definition}[Leading monomials]
For a polynomial $g(\vect)$, a monomial $\vect^{\veca} \in \supp(g)$ is said to be the \emph{leading monomial} of $g$, denoted by $\LM(g)$, if for all $\vect^{\veca'} \in \supp(g) $ we have that $\vect^{\veca'} \ltmon \vect^{\veca} $.

Similarly, we define $\LM(J) := \set{ \LM(g) : g \in J }$ for an ideal $J$.
\end{definition}

We can then define the remainder of a polynomial with respect to an ideal $J$.
\begin{definition}[Remainder modulo an ideal]
\label{defn:remainder-ideal}
For a polynomial $g(\vect)$ and an ideal $J \subset \C[\vect]$, we say that $g(\vect) \bmod J = \tilde{g}(\vect)$, if there exist polynomials $g_J(\vect) \in J$ and $\tilde{g}(\vect)$ such that $g(\vect) = g_J(\vect) + \tilde{g}(\vect)$, where $\LM(\tilde{g})$ does not belong to the ideal $\ideal{\LM(J)}$.
\end{definition}

Observe that if $LM(\tilde{g}) \not\in \ideal{\LM(J)}$, then in fact no monomial in $\supp(\tilde{g})$ belongs to the ideal $\ideal{\LM(J)}$.
And thus $\supp(\tilde{g})$ is contained in the ``complement of $\ideal{\LM(J)}$'': the \emph{normal set of $J$}.
\begin{definition}[Normal set of an ideal]
\label{defn:normal-set}
For an ideal $J \in \C[t_1,\ldots,t_r]$, the \emph{normal set of $J$} is defined as $\normalset(J) := \set{ \vect^{\veca} : \veca \in \N^r, \vect^{\veca} \not\in \ideal{\LM(J)} }$.

We sometimes overload notation to denote $\normalset(J)$ as the set of exponent vectors.
That is, $\normalset(J) = \set{\veca_1,\ldots,\veca_m}$ means $\normalset(J) = \set{\vect^{\veca_1},\ldots,\vect^{\veca_m}}$.
\end{definition}

Here are some important properties of the normal set of an ideal (see e.g. \cite{MMM93}).
\begin{fact}
\label{fact:normal-set-properties}
For any ideal $J$, its normal set $\normalset(J)$ has the following properties.
\begin{itemize} \itemsep 0pt
  \item For any $g(\vect)$, the polynomial $g \bmod J$ is a linear combination of monomials in $\normalset(J)$, and further, $\abs{\normalset(J)} = \dim_{\C}\inparen{\sfrac{\C[\vect]}{J}} $.
  \item $\normalset(J)$ is closed under divisions. That is, if $\vect^\veca \in \normalset(J)$ and $\vect^{\veca'} | \vect^{\veca} $, then $\vect^{\veca'} \in \normalset(J) $.
  In particular, $1 \in N_J $ for all ideals $J$.
\end{itemize}
\end{fact}

We can now state the result of {\Moller} and Stetter~\cite{MS95} that gives a more explicit version of the correspondence in \autoref{lem:MMM-ideal-derivative-correspondence-weak}. The following is a multivariate analogue of \autoref{fact:derivatives-characterise-quotients}.

\begin{lemma}[Consequence of {\cite[Theorem 1]{MS95}}]
\label{lem:dual-basis-ideals}
Suppose $J \subset \C[t_1,\ldots,t_r]$ is an ideal with variety $\V(J) = \set{\bar{\alpha}_1,\ldots,\bar{\alpha}_z}$ and normal set $N_J := \normalset(J) = \set{\veca_1, \ldots, \veca_w}$.
Let $\Delta_1,\ldots,\Delta_z $ be the characterising derivative operator spaces, with each $\Delta_u $ spanned by $\set{D_{u,1},\ldots,D_{u,m_u}}$, such that $\abs{N_J} = m = \sum_{u} m_u $.

Then there exists a set of $m^{2}$ constants $\set{\gamma^{(\veca)}_{u,v}} \subset \C$, such that for any polynomial $g(\vect) \in \C[\vect]$ and $\tilde{g}(\vect) := (g(\vect) \bmod J)$, we have $\coeff_{\veca}(\tilde{g}) = \sum_{u,v} \gamma^{(\veca)}_{u,v} (D_{u,v}(g))(\bar{\alpha}_u) $ for all $\veca \in N_J$. 
\end{lemma}

\section{Proof of the main theorem}
\label{sec:proof-of-theorem}

We start with an observation about diagonal ROABPs that gives an \emph{equivalent} alternate view of the model, which will be useful for our results.
\begin{observation}[Alternate view of diagonal ROABPs]
\label{obs:diagonal-roabp-alt-view}
If $f(x_1,\ldots,x_n)$ has a \emph{diagonal ROABP} of width $w$, then there is a polynomial $g(t,\vecx)$ with $\deg_t(g) \leq nw $, such that $f(\vecx) = \sum_{j \in [w]} g(j,\vecx)$.
\end{observation}
\begin{proof}
Suppose $f(\vecx) = \sum_{j \in [w]} \prod_{i \in [n]} f_{j,i}(x_i) $.
Then we define polynomials $L_1(t),\ldots,L_w(t) $ such that for each $j,k \in [w]$, $L_j(k) = 1$ if $j=k$ and $L_j(k) = 0 $ otherwise.
Such polynomials always exist, and are called \emph{Lagrange basis polynomials}.

For each $i \in [n]$, define $g_i(t,x_i) := \sum_{j \in [w]} L_j \cdot f_{j,i}(x_i) $, and let $g(t,\vecx) = \prod_{i \in [n]} g_i(t,x_i)$.
Then $g(t=j,\vecx) = \prod_{i \in [n]} f_{j,i}(x_i)$, and hence $f(\vecx) = \sum_{j \in [w]} g(j,\vecx) $ as required.
\end{proof}

\subsection{An alternate view of commutative ROABPs}

\begin{definition}
\label{defn:quotient-ring-coefficients}
For an ideal $J \subset \C[\vect]$, and a $G \in \C[\vect,\vecx]$ given by $G = \sum_{\vece} \coeff_{\vecx^{\vece}}(G)(\vect) \cdot \vecx^{\vece} $, we define the polynomial $\tilde{G} = (G \bmod J) $ as follows.
\[
    \tilde{G} := \sum_{\vece} \inparen{\coeff_{\vecx^{\vece}}(G)(\vect) \bmod J} \cdot \vecx^{\vece}
\]
Here $(g(\vect) \bmod J)$ for any $g(\vect)$ is defined as per \autoref{defn:remainder-ideal}.
\end{definition}

Using the above definition, given any commutative ROABP, we can come up with a product of univariates over $\vecx$s that is related to it in the following sense.
\begin{lemma}
\label{lem:comm-roabp-ideal-form}
Suppose $f(\vecx) = \trans{\vecb} \inparen{ \prod_{i \in [n]} \inparen{A_{i,0} + A_{i,1} x_i + \cdots + A_{i,d} x_i^d } }\vecc$, is a commutative-ROABP of width $w$ computing $f(\vecx)$.

Then there exists an ideal $J \subset \C[t_1,\ldots,t_r]$ with a finite variety, and $ G(\vect,\vecx) := \prod_i G_i(\vect,x_i)$, such that for $\tilde{G}(\vect,\vecx) := G(\vect,\vecx) \bmod J$, $f(\vecx)$ can be expressed as a linear combination of the $\vect$-coefficients of $\tilde{G}$.

Furthermore, $\abs{\vect} = r \leq \min\set{w^2,n(d+1)}$ and the $\vect$-degree of each $G_{i}$ is at most $w^2 $.
\end{lemma}

\begin{proof}
Let $F(\vecx)$ denote the $w \times w$ matrix with entries in $\C[\vecx]$, so that $f(\vecx) = \trans{\vecb} F(\vecx) \vecc$.
Let $\vecA = \set{A_1,\ldots,A_r}$ be such that the ring $\C[A_1,\ldots,A_r]$ is the same as that generated by the coefficient matrices $\set{A_{i,j}}$.
It is easy to see that $r \leq \min\set{w^2,n(d+1)}$.

We define the ideal $J$ as follows: $J = \set{ g(\vect) \in \C[\vect] : g(A_1,\ldots,A_r)=0 }$.
Let $N_{J} = \set{\vect^{\veca_1}, \ldots, \vect^{\veca_m} } $ be the \emph{normal set} of $J$; then $\abs{N_J} = m \leq w^{2}$, as the quotient ring of $J$ is isomorphic to $\C[\vecA] \subset \C^{w \times w} $ (see \autoref{lem:quotient-ring-ideal-dependencies}).
For each $i,j$  let $G_{i,j}(\vect)$ be the polynomial with monomials from $N_J $ such that $G_{i,j}(\vecA) = A_{i,j} $.
We define $G_i(\vect,x_i) = \sum_{j} G_{i,0}(\vect) x_i^j $ for each $i \in [n]$.
Since $N_J $ is closed under divisions, the degree of any $\vect^{\veca} \in N_J $ is at most $w^{2}$, and hence $\deg_{\vect}(G_i) = \deg(G_{i,j}) \leq w^2 $ for all $i$.

Let $\tilde{G} := \inparen{G \bmod J} = \sum_{\veca \in N_J} \tilde{g}_{\veca}(\vecx) \vect^{\veca}$ for some $\tilde{g}_{\veca}(\vecx) $s, which we call the ``$\vect$-coefficients of $G$''.
\begin{align*}
  f(\vecx)            &= \sum_{k,\ell \in [w]} b_k c_{\ell} \cdot F(\vecx)[k,\ell]\\
  \text{(By definition of $G$)} &= \sum_{k,\ell \in [w]} b_k c_{\ell} \cdot \inparen{G(\vecA,\vecx)}[k,\ell]\\
  \text{(By definition of $J$)} &= \sum_{k,\ell \in [w]} b_k c_{\ell} \cdot \inparen{\tilde{G}(\vecA,\vecx)}[k,\ell]\\
  \text{(Expanding $\tilde{G}$)} &= \sum_{k,\ell \in [w]} b_k c_{\ell} \cdot \inparen{\sum_{\veca \in N_J} \tilde{g}_{\veca}(\vecx) \vecA^{\veca}} [k,\ell]\\
  \text{(For $A_{\veca} = \vecA^{\veca}$)} &= \sum_{\veca \in N_J} \inparen{\sum_{k,\ell \in [w]} b_k c_{\ell} A_{\veca}[k,\ell]} \tilde{g}_{\veca}(\vecx) = \sum_{\veca \in N_J} \beta_{\veca} \tilde{g}_{\veca}(\vecx)
\end{align*}
In the last line, $A_{\veca} $ is a $w \times w$ matrix that is equal to the ``monomial'' $\vecA^{\veca} $.
\end{proof}

\subsection{Evaluating derivatives of polynomials}
We now show that for any polynomials $g(\vect)$, $h(\vect)$, and any point $\bar{\alpha} \in \C^r $, the value $(D_h(g))(\bar{\alpha})$ can be obtained as a linear combination of $O(d',\waring(h))$ evaluations of the polynomial $g$, where $d' = \max\{\deg(g),\deg(h)\}$. This is a known fact(see e.g. \cite{P19}).

\noindent We start with a fact about the ``symmetry'' between $D_h(g)(\bar{0}) $ and $D_g(h)(\bar{0}) $ that we will need.
\begin{fact}
\label{obs:symmetry-functionals}
For any $g, h \in \C[t_1,\ldots,t_r]$, $D_g(h)(\bar{0}) = D_h(g)(\bar{0}) = \sum_{\vece \in \N^r} \vece! g_{\vece} h_{\vece} $.
\end{fact}

\begin{lemma}[Functionals and Waring rank]
\label{lem:functional-waring-rank}
Let $g, h \in \C[t_1,\ldots,t_r]$ be polynomials of degree at most $d'$, and suppose $\waring(h) \leq s$.
Then there exist $W = O(s \cdot d')$ points $\vecy_{1},\ldots,\vecy_{W}$ such that $D_h(g)(\bar{0}) = D_g(h)(\bar{0}) $ can be expressed as a linear combination of $g(\vecy_1),\ldots,g(\vecy_W)$.
\end{lemma}
\begin{proof}
Let us start by expressing both $g$ and $h$ as the sum of their homogeneous components as $g = \sum_{0 \leq j \leq d'} g_j $ and $h = \sum_{0 \leq j \leq d'} h_j $.
We can therefore simplify $D_h(g)(\bar{0}) $ as follows.
\begin{align*}
  D_h(g)(\bar{0}) &= \sum_{0 \leq j \leq d'} \sum_{0 \leq j' \leq d'} D_{h_j}(g_{j'})(\bar{0}) \\
  \inparen{\text{if $j' < j$, then $D_{h_j}(g_{j'})=0$}} &= \sum_{0 \leq j \leq d'} \sum_{j \leq j' \leq d'} D_{h_j}(g_{j'})(\bar{0}) \\
  \inparen{\text{if $j' > j$, then $D_{h_j}(g_{j'})(\bar{0})=0$}} &= \sum_{0 \leq j \leq d'} D_{h_j}(g_{j})(\bar{0}) = \sum_{0 \leq j \leq d'} D_{g_j}(h_{j})(\bar{0})
\end{align*}

Now suppose that $h = \sum_{k \in [s]} (\dotProd{\vecc_k}{\vect} + b_k)^{d_k} $ is the \emph{Waring decomposition} of $h$, and let $g_j = \sum_{\vece} \coeff_{\vece}(g_j) \cdot \vect^{\vece} $, for each $j$.
We then have the following.

\begin{align*}
  D_{g_j}(h_{j})(\bar{0}) &= D_{g_j}\inparen{\sum_{k \in [s]} (\dotProd{\vecc_k}{\vect} + b_k)^{d_k}}\\
  \inparen{\text{binomial expansion to extract } h_j} &= D_{g_j}\inparen{\sum_{k \in [s]} \binom{d_k}{j} b_k^{d_k - j} \dotProd{\vecc_k}{\vect}^j }\\
  \inparen{\text{linearity of differentiation}} &= \sum_{k \in [s]} \binom{d_k}{j} b_k^{d_k - j} \cdot D_{g_j}\inparen{\dotProd{\vecc_k}{\vect}^j }(\bar{0})\\
  \inparen{\text{expanding $g_j $}}   &= \sum_{k \in [s]} \binom{d_k}{j} b_k^{d_k - j} \sum_{\vece \in \N^r} \coeff_{\vece}(g_j) \cdot \partial_{\vece} \inparen{\dotProd{\vecc_k}{\vect}^j }(\bar{0})\\
  \inparen{\text{since }\partial_{\vece} \inparen{\dotProd{\vecc_k}{\vect}^j } = j! \vecc_k^{\vece}} &= \sum_{k \in [s]} \binom{d_k}{j} b_k^{d_k - j} \sum_{\vece \in \N^r } \coeff_{\vece}(g_j) \cdot j! \cdot \vecc_k^{\vece}\\
  \inparen{\text{for $\gamma_{j,k} = \binom{d_k}{j} b_k^{d_k - j} \cdot j!$}} &= \sum_{k \in [s]} \gamma_{j,k} \cdot g_j(\vecc_k)
\end{align*}

Using previous calculations we then get that $D_h(g)(\bar{0}) = \sum_{0 \leq j \leq d'} \sum_{k \in [s]} \gamma_{j,k} \cdot g_j(\vecc_k)$.
However, this is not quite a linear combination of evaluations of $g$, because we need to ``scale'' the evaluations of $g_j $ differently for each $0 \leq j \leq d'$.
This can be easily handled using interpolation (see \autoref{lem:interpolation}) as follows, thus finishing the proof.
\begin{align*}
  D_h(g)(\bar{0}) &= \sum_{0 \leq j \leq d'} \sum_{k \in [s]} \gamma_{j,k} \cdot g_j(\vecc_k)\\
  \inparen{\text{using \autoref{cor:interpolation-homogeneous}}} &= \sum_{0 \leq j \leq d'} \sum_{k \in [s]} \gamma_{j,k} \cdot \inparen{\sum_{0 \leq \ell \leq d'} \beta_{j,\ell} \cdot g(\mu_{\ell} \vecc_k)}\\
  \inparen{\text{rearranging}}  &= \sum_{k \in [s]} \sum_{0 \leq \ell \leq d'} \inparen{ \sum_{0 \leq j \leq d'} \gamma_{j,k} \cdot \beta_{j,\ell}} g(\mu_{\ell} \vecc_k)\\
  \inparen{\text{for the appropriate $\delta_{j,\ell} $s}} &= \sum_{k \in [s]} \sum_{0 \leq \ell \leq d'} \delta_{j,\ell} \cdot g(\mu_{\ell} \vecc_k) \qedhere
\end{align*}
\end{proof}




\subsection{The proof}

We now have all the pieces required to prove the main theorem, which we first restate.

\CommROABPToDiagROABP*

\begin{proof}

  Let $F(\vecx) = \prod_{i = 1}^n \inparen{ \sum_{j = 0}^d A_{i,j} x_i^j } $, and let $f(\vecx) = \trans{\vecb} F(\vecx) \vecc$ be the corresponding commutative ROABP of width $w$.

  \begin{description}

    \item[Moving to the polynomial world:] From \autoref{lem:comm-roabp-ideal-form}, there is a $G(\vect,\vecx) = \prod_{i \in [n]} G_i(\vect,x_i) $ such that $f(\vecx)$ is a linear combination of the $\vect$-coefficients of $\tilde{G} := G \bmod J$, where $J$ is the \emph{ideal of dependencies} of the coefficient matrices $\set{A_{i,j}}$.

    Let $r = \abs{\vect}$, $\V(J) = \set{\bar{\alpha}_1,\ldots,\bar{\alpha}_z}$, and $N_J = \normalset(J) $ with $m = \abs{N_J}$.
    Then $r, m \leq w^2 $ and $\deg_{\vect}(G_i) \leq w^2 $ for all $i \in [n]$, and there exist $\beta_{\veca}$'s and $\tilde{g}_{\veca}(\vecx)$'s such that
    \[
        f(\vecx) = \sum_{\veca \in N_J} \beta_{\veca} \tilde{g}_{\veca}(\vecx).
    \]

    \item[Coefficients from derivatives:] Next, the results from \cite{MMM93,MS95} (\autoref{lem:dual-basis-ideals}) imply that there exist $m$ polynomials $\set{h_{u,v}(\vect)} $ such that:
    \begin{itemize}
      \item $\partials(h_{u,v}) \leq m$ for all $h_{u,v}$, and
      \item For any $\veca \in N_J $, $\coeff_{\veca}(\tilde{G}) = \sum_{u,v} \gamma^{\veca}_{u,v} (D_{h_{u,v}}(G))(\bar{\alpha}_u) $, for some $\set{\gamma^{\veca}_{u,v}} \subset \C$.
    \end{itemize}

    \item[Derivatives using evaluations:] Then, using \autoref{lem:functional-waring-rank} we see that for any polynomial $h$ with $s := \waring(h)$ and for any polynomial $G$ with $\deg(g),\deg(h) \leq d'$, there exist at most $s \cdot d'$ points $\vecy_1,\ldots,\vecy_{sd'} \in \C^r $ and constants $\lambda_1,\ldots,\lambda_{sd'} \in \C$ such that :
    \[
        (D_{h}(G))(\bar{\alpha}) = \sum_{q = 1}^{sd'} \lambda_q G(\vecy_q).
    \]

    Thus, for all $u,v$, $O(\waring(h_{u,v}) \cdot \max\set{\deg_{\vect}(G),\deg(h_{u,v})}) = O(S(r,m) \cdot nw^2)$ evaluations of $G$ are enough to obtain $(D_{h_{u,v}}(G))(\bar{\alpha}_u)$.

    \item[Putting everything together:] Combining all the steps, we get the following.
    \begin{align*}
        f(\vecx) \qquad &= \sum_{\veca \in N_J} \beta_{\veca} \tilde{g}_{\veca}(\vecx) \\
                        &= \sum_{\veca \in N_J} \beta_{\veca} \sum_{u,v} \gamma^{\veca}_{u,v} (D_{h_{u,v}}(G))(\bar{\alpha}_u) \\
        \text{(Rearranging)} &= \sum_{u,v} \inparen{\sum_{\veca \in N_J} \beta_\veca \gamma^{\veca}_{u,v}} (D_{h_{u,v}}(G))(\bar{\alpha}_u) \\
        \text{(For appropriate $\beta'$s)} &= \sum_{u,v} \beta'_{u,v} (D_{h_{u,v}}(G))(\bar{\alpha}_u) \\
        \inparen{\partials(h_{u,v}) \leq m,\,\, \deg(G) \leq nw^2} &= \sum_{u,v} \beta'_{u,v} \sum_{q = 1}^{S(r,m) \cdot nw^2} \lambda_q G(\vecy_q,\vecx)\\
        \therefore f(\vecx) \qquad &= \sum_{q' = 1}^{m \cdot S(r,m) \cdot nw^2} \mu_{q'} \prod_{i \in [n]} G_i(\vecy_q,x_i)
    \end{align*}

    Thus, as $m,r \leq w^2 $, we get a diagonal ROABP for $f(\vecx)$ of width $O(w^2 \cdot S(w^2,w^2) \cdot nw^2) = O(S(w^2,w^2) \cdot n w^4)$. \qedhere
  \end{description}

\end{proof}

\begin{remark}
\label{rem:connection}
Suppose there exists an explicit polynomial $f$ that is computable by a commutative ROABP of polynomial width but any diagonal ROABP computing $f$ requires width super-polynomial in $n$.
Let $w$ be the width of the commutative ROABP, and let $J$ be the ideal of dependencies of its coefficient matrices.
By \autoref{lem:dual-basis-ideals} there exist polynomials $\{h_{u,v}(\vect)\}$ with $\abs{\vect} \leq w^2 $, such that $\partials(h_{u,v}) \leq w^2$.
But if $\waring(h_{u,v}) = \poly(w)$ for each $u,v$, then we should get a diagonal ROABP of width $\poly(w)$, which is a contradiction.
Thus, a separation between commutative and diagonal ROABPs also leads to an explicit polynomial that witnesses the separation dimension of partial derivatives and Waring rank.
\end{remark}

\section{Open questions}

Owing to the connections of subclasses of ROABPs with other well-studied models, we believe that resolving any of the questions stated in \autoref{sec:introduction} in any direction would be very interesting to the algebraic complexity community, and might even lead to new approaches for PIT of ROABPs and depth 3 powering circuits.

A specific follow-up question to our main theorem(\autoref{thm:comm-roabp-to-diag-roabp}) is that of finding an appropriate converse.
For example, is it true that if diagonal ROABPs can efficiently simulate commutative ROABPs, then dimension of partial derivatives essentially captures the Waring rank of any polynomial?
It is not clear how one would go about proving the above statement directly.
For proving the contrapositive, the main technical challenge seems to be to arrive at a candidate commutative ROABP using a polynomial that would witness the separation between dimension of partial derivatives and Waring rank.

\subsection*{Acknowledgements}

We thank Ramprasad Saptharishi for numerous insightful discussions about the various structured models, which motivated this work.
We also thank Mrinal Kumar for his helpful comments about our work which helped us in enchancing the presentation.

We thank Manoj Gopalakrishan and the organisers of \emph{Thursday Theory Lunch} at IIT Bombay for organising a talk by Debasattam Pal, where we first came across the work of {\Moller} and Stetter (1995) that essentially led to the main results in this paper.

We thank the anonymous reviewers of STACS 2022 for their valuable inputs on an earlier version of the paper.

\bibliographystyle{customurlbst/alphaurlpp}
\bibliography{masterbib/references,masterbib/crossref}

\begin{thebibliography}{GKPT17}

\bibitem[AGKS15]{AGKS15}
Manindra Agrawal, Rohit Gurjar, Arpita Korwar, and Nitin Saxena.
\newblock \href {http://dx.doi.org/10.1137/140975103} {Hitting-Sets for {ROABP}
  and Sum of Set-Multilinear Circuits}.
\newblock {\em SIAM Journal of Computing}, 44(3):669--697, 2015.
\newblock Pre-print available at \href {http://arxiv.org/abs/1406.7535}
  {\path{arXiv:1406.7535}}.

\bibitem[BS83]{BS83}
Walter Baur and Volker Strassen.
\newblock \href {http://dx.doi.org/10.1016/0304-3975(83)90110-X} {The
  Complexity of Partial Derivatives}.
\newblock {\em Theoretical Computer Science}, 22:317--330, 1983.

\bibitem[BS21]{BS21}
Pranav Bisht and Nitin Saxena.
\newblock \href {http://dx.doi.org/10.1007/s00037-021-00209-y} {Blackbox
  identity testing for sum of special ROABPs and its border class}.
\newblock {\em Computational Complexity}, 30(1):8, 2021.

\bibitem[CKSV20]{CKSV20}
Prerona Chatterjee, Mrinal Kumar, Adrian She, and Ben~Lee Volk.
\newblock \href {http://dx.doi.org/10.4230/LIPIcs.CCC.2020.2} {A Quadratic
  Lower Bound for Algebraic Branching Programs}.
\newblock In {\em 35th Computational Complexity Conference, {CCC} 2020, July
  28-31, 2020, Saarbr{\"{u}}cken, Germany (Virtual Conference)}, volume 169 of
  {\em LIPIcs}, pages 2:1--2:21. Schloss Dagstuhl - Leibniz-Zentrum f{\"{u}}r
  Informatik, 2020.

\bibitem[CLO07]{CLO07}
David~A.\ Cox, John~B.\ Little, and Donal O'Shea.
\newblock \href {http://dx.doi.org/10.1007/978-0-387-35651-8} {{\em Ideals,
  Varieties and Algorithms}}.
\newblock Undergraduate texts in mathematics. Springer, 2007.

\bibitem[DDS21]{DDS21}
Pranjal Dutta, Prateek Dwivedi, and Nitin Saxena.
\newblock \href
  {https://www.cse.iitk.ac.in/users/nitin/papers/border-depth3.pdf}
  {{Demystifying the border of depth-3 algebraic circuits}}.
\newblock In {\em \FOCS{2021}}, 2021.

\bibitem[DF99]{DF99}
David~S. Dummit and Richard~M. Foote.
\newblock {\em {Abstract Algebra}}.
\newblock John Wiley and Sons, Inc., second edition, 1999.

\bibitem[DL78]{DL78}
Richard~A. DeMillo and Richard~J. Lipton.
\newblock \href {http://dx.doi.org/10.1016/0020-0190(78)90067-4} {{A
  Probabilistic Remark on Algebraic Program Testing}}.
\newblock {\em Information Processing Letters}, 7(4):193--195, 1978.

\bibitem[FGS18]{FGS18}
Michael~A. Forbes, Sumanta Ghosh, and Nitin Saxena.
\newblock \href {http://dx.doi.org/10.4230/LIPIcs.ICALP.2018.54} {Towards
  Blackbox Identity Testing of Log-Variate Circuits}.
\newblock In {\em 45th International Colloquium on Automata, Languages, and
  Programming, {ICALP} 2018, July 9-13, 2018, Prague, Czech Republic}, volume
  107 of {\em LIPIcs}, pages 54:1--54:16. Schloss Dagstuhl - Leibniz-Zentrum
  f{\"{u}}r Informatik, 2018.

\bibitem[FS13]{FS13}
\mfbiberr{toupdate(FS13): journal}Michael~A. Forbes and Amir Shpilka.
\newblock \href {http://dx.doi.org/10.1109/FOCS.2013.34} {Quasipolynomial-Time
  Identity Testing of Non-commutative and Read-Once Oblivious Algebraic
  Branching Programs}.
\newblock In {\em \FOCS{2013}}, pages 243--252, 2013.
\newblock \farXiv{1209.2408}.

\bibitem[FSS14]{FSS14}
Michael~A. Forbes, Ramprasad Saptharishi, and Amir Shpilka.
\newblock \href {http://dx.doi.org/10.1145/2591796.2591816} {Hitting sets for
  multilinear read-once algebraic branching programs, in any order}.
\newblock In {\em \STOC{2014}}, pages 867--875, 2014.

\bibitem[GG20]{GG20}
Zeyu Guo and Rohit Gurjar.
\newblock \href {http://dx.doi.org/10.4230/LIPIcs.APPROX/RANDOM.2020.4}
  {Improved Explicit Hitting-Sets for ROABPs}.
\newblock In {\em Approximation, Randomization, and Combinatorial Optimization.
  Algorithms and Techniques, {APPROX/RANDOM} 2020, August 17-19, 2020, Virtual
  Conference}, volume 176 of {\em LIPIcs}, pages 4:1--4:16. Schloss Dagstuhl -
  Leibniz-Zentrum f{\"{u}}r Informatik, 2020.

\bibitem[GKPT17]{GKPT17}
Ignacio Garc{\'{\i}}a{-}Marco, Pascal Koiran, Timoth{\'{e}}e Pecatte, and
  St{\'{e}}phan Thomass{\'{e}}.
\newblock \href {http://dx.doi.org/10.4230/LIPIcs.STACS.2017.37} {On the
  Complexity of Partial Derivatives}.
\newblock In {\em 34th Symposium on Theoretical Aspects of Computer Science,
  {STACS} 2017, March 8-11, 2017, Hannover, Germany}, volume~66 of {\em
  LIPIcs}, pages 37:1--37:13. Schloss Dagstuhl - Leibniz-Zentrum f{\"{u}}r
  Informatik, 2017.

\bibitem[GKS17]{GKS17}
Rohit Gurjar, Arpita Korwar, and Nitin Saxena.
\newblock \href {http://dx.doi.org/10.4086/toc.2017.v013a002} {Identity Testing
  for Constant-Width, and Commutative, Read-Once Oblivious ABPs}.
\newblock {\em Theory of Computing}, 13(1):1--21, 2017.
\newblock \pCCC{2016}. \arXiv{1601.08031}.

\bibitem[Kay12]{K12}
Neeraj Kayal.
\newblock \href {http://dx.doi.org/10.1145/2213977.2214036} {Affine projections
  of polynomials}.
\newblock In {\em \STOC{2012}}, pages 643--662, 2012.

\bibitem[KV21]{KV21}
Mrinal Kumar and Ben~Lee Volk.
\newblock \href {http://dx.doi.org/10.4230/LIPIcs.CCC.2021.4} {A Lower Bound on
  Determinantal Complexity}.
\newblock In {\em \CCC{2021}}, volume 200 of {\em LIPIcs}, pages 4:1--4:12.
  Schloss Dagstuhl - Leibniz-Zentrum f{\"{u}}r Informatik, 2021.
\newblock \shortECCC{20}{129}.

\bibitem[LST21]{LST21}
\mfbiberr{toupdate(LST21): everything}Nutan Limaye, Srikanth Srinivasan, and
  S{\'{e}}bastien Tavenas.
\newblock \href {https://eccc.weizmann.ac.il/report/2021/081} {Superpolynomial
  Lower Bounds Against Low-Depth Algebraic Circuits}.
\newblock {\em Electron. Colloquium Comput. Complex.}, page~81, 2021.

\bibitem[MMM93]{MMM93}
M.G. Marinari, H.M. Möller, and T.~Mora.
\newblock \href {http://dx.doi.org/10.1007/BF01386834} {Gröbner Bases Of
  Ideals Defined By Functionals With An Application To Ideals Of Projective
  Points}.
\newblock {\em Applicable Algebra in Engineering, Communication and Computing},
  4(2):103--145, 1993.

\bibitem[MS95]{MS95}
H.~Michael Möller and Hans~J. Stetter.
\newblock \href {http://dx.doi.org/10.1007/s002110050122} {Multivariate
  polynomial equations with multiple zeros solved by matrix eigenproblems}.
\newblock {\em Numerische Mathematik}, 70, 1995.

\bibitem[Nis91]{N91}
Noam Nisan.
\newblock \href {http://dx.doi.org/10.1145/103418.103462} {{Lower bounds for
  non-commutative computation}}.
\newblock In {\em \STOC{1991}}, pages 410--418, 1991.
\newblock Available on
  \href{http://citeseerx.ist.psu.edu/viewdoc/summary?doi=10.1.1.17.5067}{\tt
  citeseer:10.1.1.17.5067}.

\bibitem[NW97]{NW97}
Noam Nisan and Avi Wigderson.
\newblock \href {http://dx.doi.org/10.1007/BF01294256} {Lower bounds on
  arithmetic circuits via partial derivatives}.
\newblock {\em Computational Complexity}, 6(3):217--234, 1997.
\newblock Available on
  \href{http://citeseerx.ist.psu.edu/viewdoc/summary?doi=10.1.1.90.2644}{\tt
  citeseer:10.1.1.90.2644}.

\bibitem[Ore22]{O22}
{\O}ystein Ore.
\newblock {\"{U}}ber h{\"{o}}here Kongruenzen.
\newblock {\em Norsk Mat. Forenings Skrifter}, 1(7):15, 1922.

\bibitem[Pra19]{P19}
Kevin Pratt.
\newblock \href {http://dx.doi.org/10.1109/FOCS.2019.00053} {Waring Rank,
  Parameterized and Exact Algorithms}.
\newblock In {\em \FOCS{2019}}, pages 806--823. {IEEE} Computer Society, 2019.

\bibitem[Raz10]{Raz10}
Ran Raz.
\newblock \href {http://dx.doi.org/10.4086/toc.2010.v006a007} {Elusive
  Functions and Lower Bounds for Arithmetic Circuits}.
\newblock {\em Theory of Computing}, 6(1):135--177, 2010.

\bibitem[RS05]{RS05}
Ran Raz and Amir Shpilka.
\newblock \href {http://dx.doi.org/10.1007/s00037-005-0188-8} {Deterministic
  polynomial identity testing in non-commutative models}.
\newblock {\em Computational Complexity}, 14(1):1--19, 2005.
\newblock \pCCC{2004}.

\bibitem[Sap15]{S15}
Ramprasad Saptharishi.
\newblock \href {https://github.com/dasarpmar/lowerbounds-survey/releases/} {A
  survey of lower bounds in arithmetic circuit complexity}.
\newblock Github survey, 2015.

\bibitem[Sax08]{S08b}
Nitin Saxena.
\newblock \href {http://dx.doi.org/10.1007/978-3-540-70575-8_6} {{Diagonal
  Circuit Identity Testing and Lower Bounds}}.
\newblock In {\em \ICALP{2008}}, pages 60--71, 2008.
\newblock Pre-print available at \parseECCC{TR07/124}.

\bibitem[Sch80]{S80}
Jacob~T. Schwartz.
\newblock \href {http://dx.doi.org/10.1145/322217.322225} {{F}ast
  {P}robabilistic {A}lgorithms for {V}erification of {P}olynomial
  {I}dentities}.
\newblock {\em Journal of the ACM}, 27(4):701--717, 1980.

\bibitem[Shi16]{S16}
Yaroslav Shitov.
\newblock \href {http://arxiv.org/abs/1611.01559} {How hard is the tensor
  rank?}, 2016.
\newblock Pre-print available at \href {http://arxiv.org/abs/1611.01559}
  {\path{arXiv:1611.01559}}.

\bibitem[Str69]{S69}
V.~Strassen.
\newblock \href {http://dx.doi.org/10.1007/BF02165411} {{Gaussian Elimination
  is not Optimal}}.
\newblock {\em Numerische Mathematik}, 13(3):354--356, 1969.

\bibitem[SW01]{SW01}
Amir Shpilka and Avi Wigderson.
\newblock \href {http://dx.doi.org/10.1007/PL00001609} {Depth-3 arithmetic
  circuits over fields of characteristic zero}.
\newblock {\em Computational Complexity}, 10(1):1--27, 2001.
\newblock \pCCC{1999}.

\bibitem[Val79]{V79}
Leslie~G. Valiant.
\newblock \href {http://dx.doi.org/10.1145/800135.804419} {{Completeness
  Classes in Algebra}}.
\newblock In {\em \STOC{1979}}, pages 249--261, 1979.

\bibitem[Zip79]{Z79}
Richard Zippel.
\newblock \href {http://dx.doi.org/10.1007/3-540-09519-5_73} {Probabilistic
  algorithms for sparse polynomials}.
\newblock In {\em Symbolic and Algebraic Computation, {EUROSAM} '79, An
  International Symposiumon Symbolic and Algebraic Computation}, volume~72 of
  {\em Lecture Notes in Computer Science}, pages 216--226. Springer, 1979.

\end{thebibliography}

\appendix

\section{PIT algorithms for ROABP classes}
\label{sec:pit}

A dual question to that of proving strong lower bounds against a class $\calc$ of polynomials, is the algorithmic task of \emph{polynomial identity testing (PIT)} for $\calc$, which is as follows.
\begin{quote}
Given access to a polynomial $f \in \calc$, determine whether $f$ is identically zero.
\end{quote}

A PIT algorithm is said to be \emph{blackbox} if it is only allowed to evaluate $f$ at certain points and it is called \emph{whitebox} when the algorithm is allowed to examine how $f$ is computed in the class $\calc$ (e.g. all the entries in the matrices used by an ABP).
Note that the \emph{Polynomial Identity Lemma}~\cite{O22,DL78,Z79,S80} (also called ``Schwartz-Zippel lemma'') immediately gives an efficient \emph{randomised} blackbox PIT for all $n$-variate polynomials of degree $\poly(n)$.
Thus, designing efficient, \emph{deterministic} PITs is the interesting task here, which remains a long-standing open problem.
In fact, finding efficient deterministic PITs for well-studied classes of polynomials like $\VBP$, $\VP$, etc., is a central problem in algebraic complexity theory.

Obtaining efficient deterministic PITs even for very structured classes of polynomials turns out to be a non-trivial task.
In this section, we focus on structured ROABP classes.
Recall the following hierarchy of polynomials (where $\calc$ denotes $\calc(n,d,\poly(n,d))$).  
\[
  \SES \subsetneq \diagROABP \subseteq \commROABP \subseteq  \ROABP[\forall] \subsetneq \ROABP[\exists]
\]

Although, we have made considerable progress towards obtaining PIT algorithms, it is important to note that we do not know polynomial time deterministic blackbox PIT algorithms for any of the classes in the hierarchy mentioned above.

Firstly, for non-commutative ABPs, an efficient whitebox PIT is known due to Raz and Shpilka~\cite{RS05}, which also extends to ROABPs. Furthermore, a work of Forbes and Shpilka~\cite{FS13} provides a quasipolynomial time blackbox PIT for ROABPs \emph{in known order}, which subsumes blackbox PIT for non-commutative ABPs. Their algorithm requires the knowledge of the order in which the ROABP reads the variables; this setting is sometimes called \emph{grey-box}. This immediately gives a quasipolynomial time blackbox PIT for $\ROABP[\forall](n,d,w)$.

However, \emph{fully blackbox} quasipolynomial time PITs for ROABPs were provided by Forbes, Saptharishi and Shpilka~\cite{FSS14} ($(ndw)^{O(\log^2 n)} $) and later by Agrawal, Gurjar, Korwar and Saxena~\cite{AGKS15} ($(ndw)^{O(\log n)} $). By ``fully blackbox'', we mean that their algorithms work for the class $\ROABP[\exists](n,d,w)$ as opposed to those of \cite{FS13} which only worked for $\ROABP[\forall](n,d,w)$. A work of Gurjar, Korwar and Saxena~\cite{GKS17} gives a blackbox PIT for $\ROABP[\forall](n,d,w)$ that runs in time $n^{O(\log w)}$, which is efficient when the width $w$ is a constant. For the general case, the state of the art is a blackbox PIT given by Guo and Gurjar~\cite{GG20} which achieves the parameters of \cite{AGKS15} and improves upon them in some special cases.

Despite having an exact characterisation due to Nisan\cite{N91} that leads to nearly optimal lower bounds, obtaining efficient blackbox PITs for ROABPs remains widely open.
That said, perhaps the \emph{simplest} model for which we know of nearly optimal exponential lower bounds, but have no efficient blackbox PITs, is that of \emph{depth 3 powering circuits}.
The work of Saxena~\cite{S08b} shows that $\SES$ circuits \emph{efficiently} reduce to ROABPs (in fact $\diagROABP$), which immediately gives an efficient whitebox PIT using \cite{RS05}.
In the blackbox setting, while the previously mentioned works \cite{FS13,AGKS15} trivially extend to $\SES$, the best known blackbox PIT is due to Forbes, Saptharishi and Shpilka~\cite{FSS14}, that runs in time $n^{O(\log \log n)} $ for circuits of size and degree $\poly(n)$. In the special case of $\SES$ circuits of size and degree $s$, that depend on $O(\log s)$ variables, a recent work of Forbes, Ghosh and Saxena~\cite{FGS18} gives a blackbox PIT that runs in time $\poly(s)$.

Interestingly, the above mentioned ideas from \cite{FSS14} (alongwith the \emph{duality trick} of \cite{S08b} and the low-variate PIT by \cite{FGS18}) is known to reduce the blackbox PIT of $n$-variate, degree $d$, size $s$ $\SES$ circuits to the blackbox PIT of a diagonal ROABP of size and degree $\poly(n,d,s)$ that depend on just $O(\log(sd))$ variables!
Rather annoyingly, even then obtaining efficient blackbox PIT for $\SES$ circuits remains open. We give the exact statement here for completeness.

\begin{theorem}[{See e.g. \cite[Lemma 2.12]{BS21}}]
Let $P$ be an $n$-variate polynomial of degree $d$ computable by a size $s$ $\SES$ circuit. Then there exists a polynomial $P'$ on $O(\log sd)$ variables of degree and size $\poly(n,d,s)$ computable by a diagonal ROABP such that $P\equiv 0$ if and only if $P'\equiv 0$.
\end{theorem}

\section{Formal statements of some useful facts}
\label{sec:basics}

\begin{lemma}[Univariate interpolation (Folklore)]
\label{lem:interpolation}
Let $\mu_0,\ldots,\mu_D \in \C$ be distinct. Then there exist constants $\set{\beta_{j,k}}_{0 \leq j,k \leq D} $ such that for any polynomial $p(v) \in \C[v]$ of degree at most $D$, we have that $\coeff_{v^j}(p(v)) = \sum_{0 \leq k \leq D} \beta_{j,k} \cdot p(\mu_k) $ for all $0 \leq j \leq D$. 
\end{lemma}

\begin{corollary}[Interpolating homogeneous components]
\label{cor:interpolation-homogeneous}
For any polynomial $f(\vecx) \in \C[\vecx]$ and any $0 \leq j \leq \deg(f)$, the degree $j$ homogeneous component of $f$ denoted by $f_j $ can be expressed as a linear combination of $f(\mu_0 \cdot \vecx),\ldots,f(\mu_D \cdot \vecx)$ for any distinct $\mu_0,\ldots,\mu_D $.
\end{corollary}
\begin{proof}
Note that $f_j(\vecx) = \coeff_{v^j}(f(v \cdot x_1,\ldots, v \cdot x_n))$, and apply \autoref{lem:interpolation}.
\end{proof}

It is useful to note Nisan's characterisation for Read-once Oblivious ABPs.
Nisan~\cite{N91} showed that the width of a non-commutative ABP is exactly characterized by the rank of {\em partial derivative matrices} that we now define.
Since ROABPs are commutative analogues of non-commutative ABPs, the same characterisation extends to ROABPs, as follows.

Let $f\in \C[\vecx]$ be an $n$-variate polynomial of individual degree $< d$, and suppose we wish to compute $f$ using an ROABP in the ``sorted'' order $(x_1,x_2,\ldots,x_n)$.
For any $i \in [n]$ we then define the $d^i \times d^{n-i}$ matrix $M_i^{(f)}$ as follows.
The rows of $M_i^{(f)}$ are indexed by monomials in the first $i$ variables ($\set{x_1,\ldots,x_i}$), and its columns are indexed by monomials in the other $n-i$ variables; the entry $M_i^{(f)}[m,m']$ is the coefficient of the monomial $m\cdot m'$ in $f$.
As the entries of $M_i^f$ are in $\C$, $\rank(M_i^f)$ is well-defined.
We now state the version of Nisan's result that exactly characterises the size of the smallest ROABP computing $f$.
\begin{lemma}[Nisan's characterization for ROABPs]
\label{lem:nisan-roabp}
For any $n$-variate polynomial $f$ of degree $d$, the smallest ROABP that computes $f$ in the order $(x_1,x_2,\ldots,x_n)$ must have size exactly $\rank(M_1^{(f)}) +\cdots + \rank(M_{n-1}^{(f)})$.
\end{lemma}

\end{document}